\documentclass[]{aastex631}

\usepackage{amsmath, amssymb}
\usepackage{parskip}
\usepackage{url}

\newcommand{\br}{\mathbf{r}}
\newcommand{\boldv}{\mathbf{v}}
\newcommand{\boldB}{\mathbf{B}}
\newcommand{\bvn}{\mathbf{v_0}}
\newcommand{\bxi}{\boldsymbol{\xi}}
\newcommand{\omegat}{\tilde{\omega}}
\newcommand{\matr}[1]{\textsc{#1}}

\usepackage{xcolor}
\usepackage{soul}

\shortauthors{Brughmans, Keppens \& Goedbloed}

\begin{document}

\title{Parametric Survey of Nonaxisymmetric Accretion Disk Instabilities: Magnetorotational Instability to Super-Alfvénic Rotational Instability}

\author[0000-0002-7885-4554]{Nicolas Brughmans}
\affiliation{Centre for mathematical Plasma-Astrophysics, Celestijnenlaan 200B, 3001 Leuven, Belgium.}

\author[0000-0003-3544-2733]{Rony Keppens}
\affiliation{Centre for mathematical Plasma-Astrophysics, Celestijnenlaan 200B, 3001 Leuven, Belgium.}

\author[0000-0002-8794-472X]{Hans Goedbloed}
\affiliation{DIFFER - Dutch Institute for Fundamental Energy Research, De Zaale 20, 5612 AJ Eindhoven, the Netherlands.}

\begin{abstract}
    Accretion disks are highly unstable to magnetic instabilities driven by shear flow, where classically, the axisymmetric, weak-field Magneto-Rotational Instability (MRI) has received much attention through local WKB approximations. In contrast, discrete non-axisymmetric counterparts require a more involved analysis through a full global approach to deal with the influence of the nearby magnetohydrodynamic (MHD) continua. Recently, rigorous MHD spectroscopy identified a new type of an ultra-localised, non-axisymmetric instability in global disks with super-Alfvénic flow. These Super-Alfvénic Rotational Instabilities (SARIs) fill vast unstable regions in the complex eigenfrequency plane with (near-eigen)modes that corotate at the local Doppler velocity and are radially localised between Alfvénic resonances. Unlike discrete modes, they are utterly insensitive to the radial disk boundaries. In this work, we independently confirm the existence of these unprecedented modes using our novel spectral MHD code \texttt{Legolas}, reproducing and extending our earlier study with detailed eigenspectra and eigenfunctions. We calculate growth rates of SARIs and MRI in a variety of disk equilibria, highlighting the impact of field strength and orientation, and find correspondence with analytical predictions for thin, weakly magnetised disks. We show that non-axisymmetric modes can significantly extend instability regimes at high mode numbers, with maximal growth rates comparable to those of the MRI. Furthermore, we explicitly show a region filled with quasi-modes whose eigenfunctions are extremely localised in all directions. These modes must be ubiquitous in accretion disks, and play a role in local shearing box simulations. Finally, we revisit recent dispersion relations in the Appendix, highlighting their relation to our global framework.
\end{abstract}

\section{Introduction}

Accretion disks are generally assumed to be in a turbulent state, as observed accretion rates are higher than expected from angular momentum transport through molecular viscosity alone \citep{pringle1981,BH98}. The origin of this turbulence and the resulting enhanced viscosity has long been unclear: hydrodynamic turbulence is largely excluded in typical accretion disk flow profiles that satisfy the Rayleigh criterion for linear stability. Non-axisymmetric hydrodynamic instability pathways, both linear and non-linear, for  differentially rotating tori were identified early on \citep{papaloizou-pringle, blaeshawley}, but these more global modes are generally not expected to lead to fully developed turbulence in disks.

Only some thirty years ago, a convincing linear route to turbulence in disks was discovered through the weak-field, axisymmetric Magneto-Rotational Instability (MRI) \citep{BH91}, revisiting earlier analysis by \citet{velikhov1959} and \citet{chandrasekhar1960}. Even a seed vertical magnetic field was shown to dramatically alter stability by acting as a tether between fluid elements, transporting angular momentum outward. Non-linear simulations confirmed how, at unstable MRI wavelengths, the quickly developing field distortions lead to angular momentum interchanges \citep{HB91}. \cite{HB92} extended these non-linear 2D simulations in a shearing sheet to address their long-term evolution and found magnetic energy growth and sustained angular momentum transport. The linearly unstable, quickly growing MHD eigenmodes induce the generation of turbulent eddies at multiple scales, whose interactions may explain enhanced effective viscosity through the stress tensor. In the meantime, MRI has become an accepted pathway to disk turbulence through linear MHD instability in many fully 3D non-linear MHD simulations, from fully global Newtonian \citep{mishra2019} and general relativistic settings \citep{begelman22, ripperda2022} to local 3D shearing boxes in ideal or resistive MHD \citep{lesaffre2009,bai_stone13,hirai2018} or even kinetic particle-in-cell settings \citep{bacchini2022,bacchini2024}.

The linear dispersion relation for the standard weak-field MRI is obtained from a WKB analysis in a Cartesian shearing sheet or box, where the background curvature is ignored \citep{BH91}. This local approach is in perfect correspondence with results from global cylindrical studies of the MRI where the full background variations are taken into account \citep{blokland05, latter15}. However, the linear MHD spectrum of eigenoscillations is intricately organised and truly sensitive to background variations, as exposed by MHD spectroscopy, which has its theoretical foundations in laboratory plasma contexts \citep{GKP19}. Early examples of MHD spectroscopy in cylindrical models relate to rotating screw-pinches \citep{hameiri1981}, but it has also been applied to accretion disk theory \citep{keppens2002} to obtain the general equations governing all linear eigenmodes of any gravitating, rotating, magnetised disk with radial variation. 

In contrast to the axisymmetric MRI, non-axisymmetric modes fundamentally require such a global spectroscopic analysis because of singularities in the governing equations produced by the overlap of the continuous MHD spectra for radially varying, rotating equilibrium states. These non-axisymmetric instabilities are of particular interest for dynamo activity, which is impossible under axisymmetric conditions \citep{moffatt1978}, and because their coupling to the toroidal field is essential for self-sustained instability \citep{hawley1996,kitchatinov2010,begelman22}. Furthermore, although global, they can become very localised with growth rates as high as those of the axisymmetric MRI, and may still be unstable under conditions where the MRI is typically stabilised \citep{GK22, begelman22}, as is also shown in this work. Additionally, secondary parasitic non-axisymmetric instabilities developing on channel flow solutions in shearing boxes have growth rates that can even exceed that of the MRI \citep{goodmanxu94}. Although the original MRI description was immediately followed by a local WKB description of the non-axisymmetric MRI featuring transient growth \citep{BH92b}, a systematic MHD spectroscopic approach has shown the existence of unstable modes that cannot be found from local theory alone \citep{matsumoto_tajima95, ogilvie_pringle, GK22}. Additionally, growth rates for these `local' non-axisymmetric MRI are an order of magnitude underestimated as compared to those of their `global' versions \citep{curry_pudritz96}.

Research on the global linear treatment of non-axisymmetric modes peaked in the decade after the MRI was discovered: \citet{matsumoto_tajima95} performed a global analysis of a local shearing sheet and found evanescent modes localised between two Alfvén resonances; \citet{curry_pudritz96} investigated a cylindrical equilibrium in the Boussinesq approximation with a constant vertical field and found two corotating Alfvénic modes, generalising the hydrodynamic Papaloizou-Pringle instability \citep{papaloizou-pringle}; \citet{terquem_papaloizou} performed linear analysis and simulations of a cylindrical disk having both radial and vertical structure and limited to a toroidal field, and in their simulations obtained unstable modes growing on increasingly smaller scales; \citet{ogilvie_pringle} compared non-axisymmetric modes in both a cylinder and a shearing sheet for a purely toroidal field, and found two branches of modes clustering towards the continua in the limit $k\rightarrow+\infty$ with arbitrarily localised eigenfunctions; \citet{noguchi2000} further explored the evanescent eigenmodes found by \citet{matsumoto_tajima95}, although the range of unstable fields in their shearing sheet is at odds with the earlier cylindrical models; and \citet{keppens2002} gave a fully general theoretical and numerical description of the MHD eigenspectrum of a cylindrical model and found an abundance of unstable non-axisymmetric modes. This general formalism has yet to be applied to perform a comprehensive linear study of non-axisymmetric accretion disk instabilities, which is what we aim to accomplish in this work. In contrast to many of the above works, we will consider an arbitrary magnetic field orientation and strength to show how the predictions from these limiting cases (purely vertical and azimuthal fields) hold up in general.

In recent years, non-axisymmetric mode analysis has been sporadically revived, mostly in local shearing box models where the radial wavenumber changes in time and the instability is transient \citep{kitchatinov2010,mamatsashvili13} or in WKB and global treatments of cylindrical models with a dominant azimuthal field \citep{das18,begelman22} as it becomes clear that these modes must play an important role in the dynamics of accretion disks. However, in non-linear simulations, the development of turbulence is still often attributed solely to the axisymmetric MRI. Concurrently, there is claimed laboratory evidence for non-axisymmetric MRI in an azimuthal field \citep{seilmayer2014}, for the standard vertical-field MRI \citep{wang2022sMRI}, and for unidentified non-axisymmetric modes \citep{wang2022non-axisymm}, to which linear global analysis has also been applied \citep{EP22}. All this proves that these are still exciting times for linear studies of non-axisymmetric instabilities, contrary to popular belief. 

Indeed, \citet{GK22} recently found a completely new type of linear ideal MHD instability in thin, weakly magnetised, cylindrical accretion disks that could provide a broadly accessible pathway to turbulence. These modes densely populate a 2D region in the complex eigenfrequency plane and have eigenfunctions that resemble radially localised wave packages. \added{The instability is termed Super-Alfvénic Rotational Instability (SARI), which reflects the fact that a disk has rotational Doppler frequency $m v_\theta/r$ dominating over the Alfvén frequency if it is assumed thin and weakly magnetised.} These modes are promising because of the apparent ease with which they can be excited (they are not normal modes but require only a tiny addition of energy to be brought into resonance), their insensitivity to boundary conditions (the eigenfunctions are radially bound by Alfvén resonances) and the property that they exist all over the disk (each perturbation peaking at the local Doppler corotation radius). SARIs exist because of a delicate interplay between the MHD continua and the (near-)singularities they introduce in the analysis of the governing ordinary differential equations, hence they can only be obtained through a global analysis. \citet{GK22} provided full analytic as well as numerical evidence for SARIs, using the powerful Spectral Web approach \citep{goedbloed2018spectralI,goedbloed2018spectralII} to locate complex eigenfrequencies of ideal, stationary MHD states. In addition, they rigorously generalised a criterion for instability and growth rate prediction. In this work, we aim to confirm the existence of SARIs with an independent numerical approach and to show the applicability of this instability criterion.

To distinguish between the transient growth of local non-axisymmetric MRI as described in the literature \citep{BH92b, kitchatinov2010,mamatsashvili13} and the global eigenmode nature of the non-axisymmetric SARIs, we will in the remainder of this work refer to any axisymmetric mode as \textit{the MRI}, and to non-axisymmetric modes as SARIs in the regime of thin, weakly magnetised disks. We address some properties of non-axisymmetric modes that are known from local theory or limiting cases in global analysis but are underexposed in the general global approach. In particular, we confirm analytical predictions for instability over a range of equilibria and show that the Alfvén frequency plays a central role, which opens up more regimes to instability. Furthermore, we show that global high-$m$ modes are nevertheless very localised, with possible implications for shearing box simulations.

In this work, we first revisit, augment, and further validate the results of \citet{GK22} using the \texttt{Legolas} code \citep{claes2020legolas}. The equilibrium set-up and numerical approach are described in Sect.~\ref{sec:methods}. Section~\ref{sec:MRI_SARI} summarises the properties of the novel quasi-continuum modes, as well as those of the standard MRI and discrete SARI modes in a global approach. It will become clear that resonances with the Alfvén continua play a crucial role in the behaviour of non-axisymmetric modes. Second, we perform a comprehensive study of the most unstable MRI and SARI growth rates in a range of equilibria. Sections~\ref{sec:vertical_localisation}--\ref{sec:magnetisation} compare growth rates for the MRI and SARIs and show that non-axisymmetric modes can significantly extend regimes of instability to stronger fields. These results are shown to match classical and new analytical predictions in Sect.~\ref{sec:instability_criterion}, and a comparison with local dispersion relation solutions highlights the necessity of a global approach. Finally, we again focus on the full MHD eigenspectra in Sect.~\ref{sec:param_spectra} to explicitly show a quasi-continuum at a high azimuthal wavenumber, featuring an abundance of modes with significant growth rates that are global in nature but radially and vertically extremely localised. As additional material, Appendix~\ref{sec:appendix_spectral_web} summarises the mathematical tools required for a global and local analysis. Appendix~\ref{sec:appendix_dispersion_blokland} compares the general local dispersion relation of \citet{blokland05} and other recent dispersion relations that have been used for non-axisymmetric modes \citep{das18,EP22}.

\section{Methods} \label{sec:methods}
\subsection{Model equilibrium} \label{sec:model_equilibrium}

Our basic model representing a cylindrical accretion disk is an ideal MHD equilibrium \citep{keppens2002,blokland05,GK22} that radially varies according to self-similar profiles \citep{spruit1987} and that is vertically invariant. The disk is modelled as an annulus with radius $r \in [r_1, r_2]$, with a line source approximating the central gravitating object, $\mathbf{g}(\br) \approx \frac{G M_*}{r^2} \mathbf{\hat{r}}$. The assumption of vertical invariance in the disk is valid if small vertical wavelengths (or large wavenumbers $k$) are considered, much smaller than the scale height $H = c_S/ (\Omega_1 r_1) \ll r_1$ in a thin, Keplerian disk \citep{shakura1973}, where $c_S = \sqrt{\gamma p/\rho}$ is the sound speed and $\Omega_1$ the disk rotation frequency. From here on, we eliminate all proper units from the variables through  the equilibrium values at the inner boundary $r_1$, all denoted by a subscript $1$. The normalisation is fixed by choosing new units for density $\rho_1$, distance $r_1$ and velocity $v_{K1}=\sqrt{{GM_*}/{r_1}}$ (the inner Keplerian rotation velocity). The non-dimensionalised radial equilibrium profiles for the density, pressure, velocity, and magnetic field are then given by
\begin{equation} \label{eq:equilibria}
    \rho(r) = r^{-3/2}, \quad p(r) = p_1 r^{-5/2}, \quad \boldv(r) = \Omega_1 r^{-1/2}\boldsymbol{\hat{\theta}}, \quad \boldB(r) = B_{\theta1} r^{-5/4} \boldsymbol{\hat{\theta}} + B_{z1} r^{-5/4}\mathbf{\hat{z}},
\end{equation}
such that the stationary ($\partial/\partial t = 0$) MHD force-balance equation is satisfied,
\begin{equation} \label{eq:radial_force_balance}
    r\Lambda(r) := \rho r \left( \Omega^2 - \frac{1}{r^3} \right) = \left( p + \frac{1}{2} B^2 \right)' + \frac{B_\theta^2}{r}.
\end{equation}
Here, $\Lambda = 0$ if the flow is Keplerian, with modifications by pressure gradients and magnetic tension. The dimensionless radial domain is $r \in [1, 1+\delta]$, where the parameter $\delta = (r_2-r_1)/r_1$ describes the size of the domain. It is instructive to introduce three additional parameters that determine the constants at the inner disk edge in Eq.~\eqref{eq:equilibria} \citep{GK22}: magnetic field strength or Alfvén speed $\epsilon := B_1 = B_1 / \sqrt{\rho(1)} = v_{A1}$, plasma-beta $\beta := 2 p_1/B_1^2$, and inverse pitch angle $\mu_1 = B_{\theta1}/B_{z1}$. It should be noted that the general inverse pitch angle $\mu := B_\theta/(r B_z)$ is actually radially dependent, but that the plasma-beta is uniform throughout the disk as $p$ and $B^2$ have the same $r$-variation. The latter also implies that $\beta = 2 c_S^2/\gamma v_A^2$ is constant, so the varying sound speed is a fixed multiple of the varying Alfvén speed. We take $\gamma = 5/3$ for a monatomic gas.

In terms of these free parameters, which fully determine an equilibrium, the constants at the inner boundary become
\begin{equation} \label{eq:inner_values}
    p_1 = \frac{1}{2} \beta\epsilon^2, \quad B_{z1} = \frac{\epsilon}{\sqrt{1+\mu_1^2}}, \quad B_{\theta1} = \mu_1 B_{z1} = \frac{\epsilon\mu_1}{\sqrt{1+\mu_1^2}}, \quad \Omega_1 = \sqrt{1 - \frac{5}{4} \epsilon^2 \left(\beta + \frac{1+\frac{1}{5}\mu_1^2}{1+\mu_1^2}\right)},
\end{equation}
where the rotation frequency is obtained from the radial force balance equation. Note that, although the rotation power-law is Keplerian, the actual rotation frequency $\Omega_1$ is in general not Keplerian because of pressure and tension forces. For convenience, we denote $\varphi = \arctan{B_{\theta1}/B_{z1}}$ for the angle between the equilibrium magnetic field and the $z$-axis. Variation of the field orientation from poloidal (vertical) to toroidal (azimuthal) is hence achieved through varying $\varphi$ from $0^\circ$ to $90^\circ$.

\subsection{Eigenvalue problem} \label{sec:eigenvalue_problem}

The force-balanced equilibrium given above can then be analysed for its eigenoscillations, i.e. the complete set of complex eigenfrequencies and eigenfunctions of linear perturbations. Perturbed quantities $f$ are assumed to be of the form
\begin{equation} \label{eq:fourier}
    f(r,\theta,z,t) = \hat{f}(r) \exp{i\left(m\theta+kz - \omega t\right)},
\end{equation}
corresponding to a Fourier decomposition in the spatial and temporal domains with complex Fourier coefficients $\hat{f}(r)$, real azimuthal and vertical wavenumbers $m$ and $k$, respectively, and complex eigenfrequencies $\omega = \sigma + i\nu$. Due to the $\theta$-periodicity of the cylindrical equilibrium, $m$ is quantised to integer values, whereas the vertical wavenumber $k$ can take arbitrary values in this case. The set of eigenoscillations of this equilibrium is fully determined with the wavenumbers $m$ and $k$.

We calculate all MHD eigenspectra in this work using the \texttt{Legolas}\footnote{\url{https://legolas.science}. The code version used in this work is open-source under the GPL-3.0 license and available on GitHub at \url{https://github.com/n-claes/legolas/tree/v2.0.0}.} code developed by \citet{claes2020legolas} and \citet{legolasjordi} and recently upgraded in terms of solver and memory performance \citep{legolas2}. It uses a finite-element approach to solve the following generalised matrix eigenvalue problem obtained from the linearised MHD equations, where the Fourier coefficients become eigenfunctions:
\begin{equation} \label{eq:legolas_eigval_problem}
    \matr{A}(r) \mathbf{X}(r) = \omega \matr{B}(r) \mathbf{X}(r).
\end{equation}
Here, $\mathbf{X}$ is an 8-component eigenvector before its Galerkin decomposition (dropping the hat on the Fourier coefficients), containing eigenfunctions $(\rho, v_r, v_\theta, v_z, T, a_r, a_\theta, a_z)$, with $\rho$ the density perturbation, $T$ the plasma temperature perturbation and $\mathbf{a}$ the magnetic vector potential, which fixes the perturbed magnetic field $\mathbf{B} = \nabla\times\mathbf{a}$ and enforces the monopole constraint $\nabla\cdot\mathbf{B}$ exactly. \texttt{Legolas} constructs the matrices $\matr{A}$ and $\matr{B}$ from the equilibrium variations, the chosen finite-element base functions in the eigenmode representations, and the required physical effects like viscosity, resistivity, Hall MHD, etc. \citep{legolasjordi}. As this work is limited to ideal MHD, the matrices $\matr{A}$ and $\matr{B}$ are modified from the static, linear MHD case with only flow and gravity as additional effects. The matrix system is then solved using various available solvers depending on the requirements. In this work, we use the QR-invert algorithm to produce complete eigenspectra, and the Arnoldi shift-invert method to search for eigenvalues in the neighbourhood of some complex number \citep{legolas2}. The mathematical formulation of the eigenvalue problem is closed by fixing the boundary conditions at radius $1$ and $1+\delta$. By default, \texttt{Legolas} assumes perfectly conducting, solid walls as boundaries, requiring $v_r=0$ and $a_\theta = a_z = 0$ at those locations. As explained in \citet{GK22}, these solid-wall boundaries will affect the detailed variation of MRIs, while SARIs are fully agnostic of these boundary conditions, an important distinction between both mode categories.

To substantiate our \texttt{Legolas} results, we combine them in several figures with results obtained with the complementary \texttt{ROC} code, which employs a shooting method \citep{goedbloed2018spectralI,goedbloed2018spectralII}. Appendix \ref{sec:appendix_spectral_web} gives a brief description of the formulation of the spectral problem for stationary, cylindrical equilibria in terms of a second order ODE, which forms the foundation for the \texttt{ROC} code ánd for the analytical results of \citet{GK22}. Key to our further discussion of the SARI are the MHD continua, which are singular frequency ranges that organise the spectrum. Because of the rotational equilibrium flow, all frequencies are Doppler-shifted according to the Doppler range:  
\begin{equation} \label{eq:doppler_range}
    \Omega_0(r) := m\Omega(r) + k v_{z}(r), \qquad \omegat := \omega - \Omega_0.
\end{equation}
Note that, because $v_{z}=0$ in our set-up, the Doppler shift vanishes for axisymmetric ($m=0$) perturbations. The four singular ranges are dubbed forward and backward Alfvén and slow continua, as the frequencies lie in a band on the real axis. The continua are given by,
\begin{align} 
    \Omega_A^\pm(r) &= \pm \omega_A + \Omega_0 , \quad  \Omega_S^\pm(r) = \pm \omega_S + \Omega_0,  \\
    \omega_A^2(r) &= \left(\frac{m}{r}B_\theta + kB_z\right)^2/\rho, \quad \omega_S^2(r) = \frac{\gamma p}{\gamma p + B^2} \omega_A^2,
\end{align}
which are the Doppler-shifted Alfvén and slow continua. From here on, $\Omega_{A,S}^\pm$ denotes the closed range of continuum frequencies expressed by $\{\Omega_{A,S}^\pm(r)\}$ in \citet{GK22}. In accretion disks with subthermal magnetic fields, these Alfvén and slow frequency ranges almost coincide because of incompressibility (if $\gamma \rightarrow +\infty$, they exactly coincide). Our numerical results do always account for compressibility effects, as we adopt a ratio of specific heats $\gamma=5/3$ for the disk plasma. Whereas $\pm \omega_A$ are always separated for $m=0$ modes like the MRI, the continua can overlap for non-axisymmetric modes. Exactly this phenomenon is responsible for many of the interesting properties of SARIs. For large values of $m$, $\Omega_0$ dominates over $\omega_A$ and the forward and backward continua overlap to a large degree.

\section{MRI and SARI} \label{sec:MRI_SARI}
\subsection{Accretion disk instabilities}

By means of the tools above, we can now perform an MHD spectroscopic study of accretion disk instabilities. We begin by revisiting the spectra typically associated with MRI and SARI as given by \citet{GK22} to calibrate our results. We find exact correspondence in terms of eigenvalues and eigenfunctions for both MRI and discrete SARIs.

\begin{figure}
    \plotone{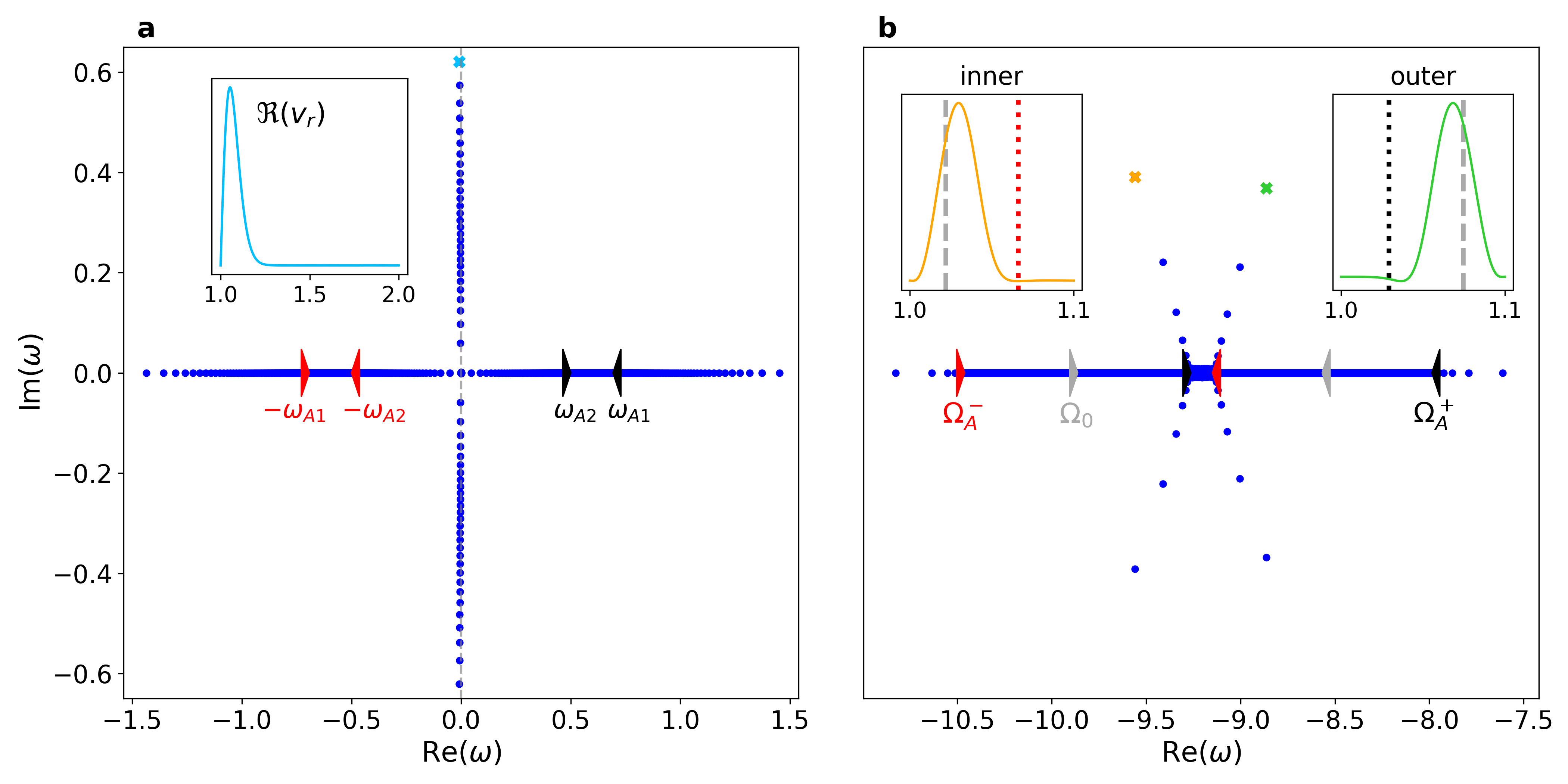}
    \caption{MRI and SARI spectra with eigenfunctions of the most unstable modes (recreation of Figs. 3 and 5 of \cite{GK22}). Note the large Doppler shift in Panel (b), where the overlapping continua produce two branches of SARIs. The corresponding eigenfunctions are localised between one boundary and an Alfvén resonance (red/black, dotted) and peak at the Doppler resonance (gray, dashed). \textit{Panel (a)}: $\delta = 1$, $\epsilon = 0.01 \sqrt{2}$, $\mu_1=1$, $\beta=100$, $m=0$ and $k=70$. Resolution of $500$ gridpoints. \textit{Panel (b)}: idem but smaller domain $\delta = 0.1$ and $m=-10$. Resolution of $200$ gridpoints.}
    \label{fig:MRI_SARI}
\end{figure}

The MHD spectrum of a typical weakly magnetised, near-Keplerian disk for axisymmetric $m=0$ modes contains the MRI, among other modes. The structure of such a spectrum is well-known \citep{keppens2002,blokland05,das18,GK22}, and is shown in Fig.~\ref{fig:MRI_SARI} (a) for a weak field with equal poloidal and toroidal components and $k=70$. Vertically, we find a finite sequence of overstable discrete MRI modes (and their damped counterparts), and horizontally, we find oscillating Alfvénic/slow modes clustering to either edge of the Alfvén continua, indicated by red and black segments on the real axis. In fact, the oscillating modes clustering to the edges $\pm\omega_{A2}$ belong to the sequence of MRI modes when listed according to the increasing number of zeroes of their eigenfunctions but they are stable (in a local MRI analog, the radial wavenumber $q$ becomes larger, stabilising the MRI). Two sequences of fast modes cluster towards infinity, but they are not shown in the figure. The up-down symmetry of the spectrum is a standard feature of ideal MHD where all operators are self-adjoint \citep{GKP19}, but the damped MRI variant is of course of limited physical significance for accretion disk turbulence (but it is significant in the initial value problem). The spectrum is, however, not left-right symmetric as a result of background flow. In particular, the axisymmetric MRI is not exponentially unstable (note the slight deviation from the imaginary axis), in contrast with local WKB predictions \citep{BH91}, although there exists a local dispersion relation that does reproduce this behaviour \citep{blokland05}. A typical eigenfunction, here the real part of $v_r$ for the most unstable eigenvalue, is shown in the inset (the imaginary part looks similar). The perturbation is localised at the inner boundary and has no zeroes inside the domain. Moving down through the sequence of MRIs, the number of zeroes increases and the perturbations become global over the entire domain. They are hence sensitive to the boundary conditions.

Although a complete spectroscopic picture of non-axisymmetric accretion disk instabilities has been missing until recently, it has been known for some time that a wavenumber $m\neq 0$ produces an additional branch of discrete unstable modes if the Alfvén continua overlap. These modes resonate with both Alfvén continua and corotate with the disk at the local Doppler frequency \citep{matsumoto_tajima95,ogilvie_pringle,curry_pudritz96}. Since a weakly magnetised, thin disk rotates super-Alfvénically ($v_{A1} \ll 1 \approx \Omega_1$), \citet{GK22} termed them super-Alfvénic rotational instabilities in their general analysis, where they showed that the two branches consist of infinitely many discrete modes clustering towards the internal edges of the Doppler-shifted continua $\Omega_A^+$ and $\Omega_A^-$. These modes are all unstable, in contrast with the finite number of unstable MRIs in Fig.~\ref{fig:MRI_SARI}~(a). Figure~\ref{fig:MRI_SARI}~(b) features a typical SARI spectrum for $m=-10$, considering a smaller domain with $\delta=0.1$ and the same equilibrium as that in Panel (a). Note the large Doppler shift that moves the oscillation frequencies to negative values, making the continua $\Omega_A^\pm$ (red and black) overlap in this super-Alfvénic regime. For these counter-rotating SARIs (since $m<0$), both $\Omega_0$ and $\Omega_A^\pm$ are decreasing, and the inner SARI eigenfrequencies cluster to $\Omega_A^-(r_1)$ whereas the outer SARIs cluster to $\Omega_A^+(r_2)$ (and vice-versa for co-rotating modes with $m>0$). The most unstable SARIs are indicated in the figure, and we note that the inner SARI growth rate is largest -- a general feature of disks with outwardly-decreasing equilibrium profiles (for an explanation, we refer to Sect.~\ref{sec:instability_criterion}). The fact that both branches have oscillation frequencies $\sigma$ inside the Doppler range (gray) implies that these modes corotate with (or against) the local flow at the Doppler corotation radius $r_*$ where $\Omega_0(r_*)=\sigma$. Since $\sigma$ is also in one of the continua $\Omega_A^\pm$, each mode has one Alfvénic resonance, which has direct bearing on the eigenfunctions shown in the two insets of Panel (b), with vertical lines denoting the resonances: the inner $v_r$ eigenfunction (orange) is localised between the inner boundary and the resonance with the backward Alfvén continuum, the outer eigenfunction (green) between the resonance with the forward Alfvén continuum and the outer boundary. Both peak at the location of resonance with the local Doppler frequency. Clearly, the names `inner' and `outer' also reflect the localisation of the eigenfunctions \citep{ogilvie_pringle}. Skin currents at the resonant location shield the mode from one boundary, acting as a virtual wall \citep{GK22}. The number of zeroes of the eigenfunctions (the `radial wavenumber') again increases down the sequence, but it is less straightforward to number them accordingly \citep{GK22}. 

\subsection{Radially extended disks}

Even though the radial domain in Fig.~\ref{fig:MRI_SARI} (b) is chosen to be small, the SARI eigenfunctions are not global because they are limited by an Alfvén resonance. In some sense, each mode actually has a second Alfvén resonance outside of the domain at a fictitious radius if the continua are extended. Intuitively, these additional resonances are encapsulated when the domain is enlarged, and the result might be hybrid modes that have properties of both inner and outer SARIs: they are localised between the two resonances, and hence shielded from the boundaries. It is clear that for a real accretion disk, these boundaries are unphysical. Physically relevant modes should hence preferentially be insensitive to these artificial constraints, like the modes in radially extended disks proposed here.

In reality, the matter is more complicated, but indeed, overlapping forward and backward shifted Alfvén continua can produce modes with two resonances, forming an unprecedented 2D region in the complex eigenfrequency plane of overstable quasi-modes, as recently reported by \citet{GK22}. In regimes with a large overlap between $\Omega_A^\pm$ (if $\delta$ is not small, or $k$ and $m$ are large), a wealth of modes appears above the Doppler range. These modes have a non-zero complementary energy $W_\text{com}$, a complex-valued quantity that should vanish for true normal modes \citep{goedbloed2018spectralI} (see Appendix \ref{sec:appendix_spectral_web}). This complementary energy quantifies the addition of energy required to excite these modes; they are almost eigenoscillations like normal modes, but not exactly. However, for these modes, $W_\text{com}$ is very small, sometimes even below machine precision, rendering the resulting near-eigenmodes numerically indistinguishable from actual normal modes -- hence \citet{GK22} termed them \textit{quasi-modes}. By allowing a small but finite $W_\text{com}$, dense regions in the complex plane containing such modes appear, from here on named continua of quasi-modes or \textit{quasi-continua}. In practice, non-zero $W_\text{com}$ implies that the total pressure perturbation shows a minute jump at the location where a shooting method links up two numerical solutions starting from the left and right boundaries, but this is numerically virtually unnoticeable at the small values of $W_\text{com} = 10^{-12}$ that we have here. Physically, $W_\text{com}$ is the additional energy required to bring the mode into resonance with the time dependency $\exp(-i\omega t)$. Such tiny excitation energies as for the quasi-modes in this work could easily be provided by ambient energy fluctuations in turbulent disks, so that there is no observable difference from a normal mode. Moreover, in actual non-linear numerical MHD simulations, one always has round-off and truncation errors, related to the spatio-temporal discretisations used. Deviations from machine precision can be sufficient to introduce non-zero complementary energy at any time during the run.

An unstable quasi-continuum has oscillation frequencies inside the Doppler range and near the Alfvén continua. An analytic treatment of the SARI quasi-modes must hence deal with three near-singular resonance locations \citep{GK22}, which are only accessible through global analysis. In \citet{GK22}, analytic and shooting-based numerical results were presented using the Spectral Web technique \citep{goedbloed2018spectralI,goedbloed2018spectralII}, emphasising the distinctive properties of these quasi-continuum SARIs. We revisit their results in the next Section using a different and independent approach with \texttt{Legolas}.

\section{Quasi-continuum modes} \label{sec:QC}

The above discussion on quasi-modes and their small but finite complementary energy represents an unusual numerical challenge to a finite-element-based solver like \texttt{Legolas}. Because of the Galerkin decomposition of eigenmodes, the boundary conditions and eigenfunction continuity (even differentiability) will be satisfied per construction, and will not show a minute jump associated with non-zero $W_\text{com}$. If a quasi-mode is found, it is \textit{a priori} truly indistinguishable from any other mode in the spectrum. However, a change in resolution will always slightly shift the quasi-mode eigenvalues found previously, whereas convergence towards a solution generally requires that an eigenvalue/eigenvector pair change minimally when increasing the resolution. This regular convergence behaviour is indeed true for discrete modes like MRI or the inner and outer, discrete SARI variants.

\begin{figure}
    \includegraphics[width=\linewidth]{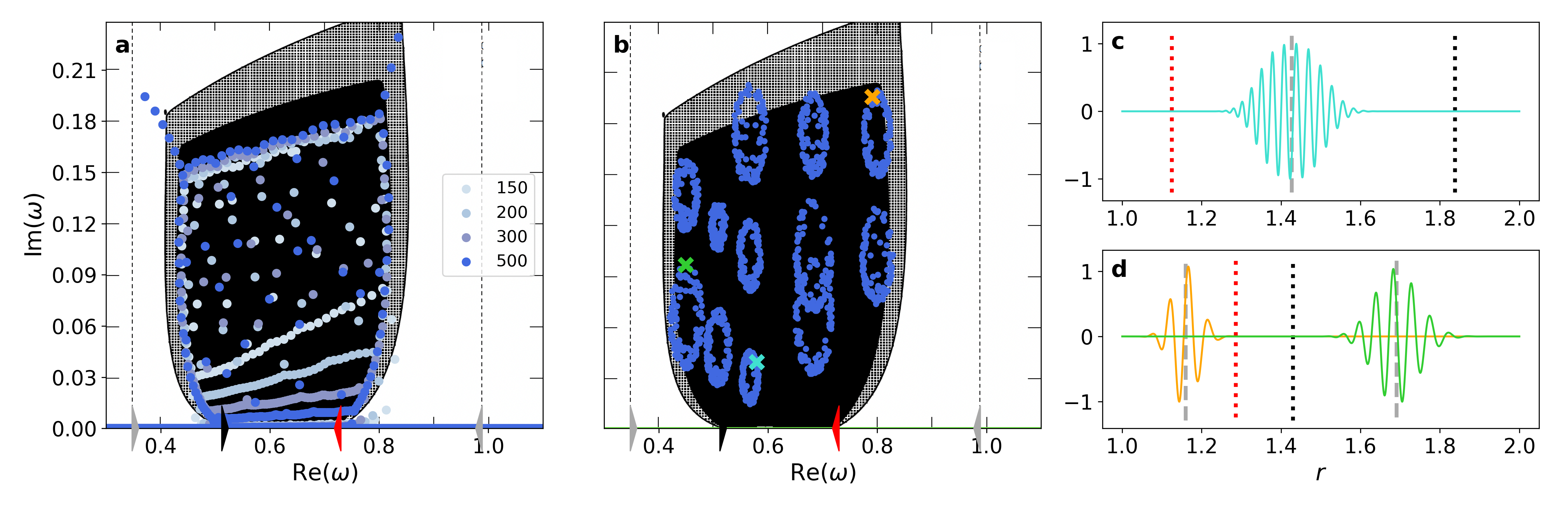}
    \caption{\texttt{Legolas} version of a quasi-continuum (Fig.~15 of \citet{GK22}), showing a correspondence with $W_\text{com} \leq 10^{-12}$ modes (black region). \textit{Panel (a):} QR-invert finds the edges of the quasi-continuum at increasing resolution. Note how the quasi-continuum bottom coincides with the overlap between $\Omega_A^-$ (red) and $\Omega_A^+$ (black). \textit{Panel (b):} shift-invert runs at $1000$ gridpoints. The annular distributions imply the presence of many nearby modes. \textit{Panels (c,d):} eigenfunctions $\Re(v_r)$ for the three quasi-modes on Panel~(b). Equilibrium parameters: $\delta = 1$, $\epsilon = 0.045$, $\mu_1=10$, $\beta=10$, $m=1$ and $k=50$. 
    }
    \label{fig:QC_resolution}
\end{figure}

Despite these difficulties, it is possible to validate the quasi-continuum results from \citet{GK22} with \texttt{Legolas} with the combined effort of multiple solvers implemented in the code. Figure~\ref{fig:QC_resolution} overlays some of our results in Fig.~15 of \citet{GK22} for the same equilibrium. Panel~(a) shows the $m=1$ co-propagating SARI spectrum found using the QR-invert algorithm for increasing resolution $n_\text{grid}\in\{150,200,300,500\}$. The spectra are overplotted on the $W_\text{com} \leq 10^{-12}$ (black) and $10^{-8}$ regions calculated with the \texttt{ROC} code and show a good match: the QR-invert algorithm locates the left and right edges of the quasi-continuum, as well as additionally finding the discrete inner and outer SARI modes clustering to the continua and disappearing in a sea of quasi-modes. For increasing grid resolution, the upper edge of the quasi-continuum as calculated by \texttt{Legolas} remains more or less stationary, but the lower edge shifts downwards, thereby increasingly resolving the lateral edges. Since the independent Spectral Web approach can identify the entire black shaded region as having a minute complementary energy, we must conclude that the lower edge of the quasi-continuum obtained by the QR algorithm is hence an artificial feature of the limited resolution. The correspondence between the quasi-continuum edges and the black region for $W_\text{com} = 10^{-12}$ (small compared to the normalised eigenfunctions) indicates that \texttt{Legolas} operates on a similar accuracy. It should be noted that low-resolution versions of Fig.~\ref{fig:MRI_SARI}~(b) also show a quasi-continuum like structure between the two SARI branches, but it becomes smaller as the resolution is increased and its upper edge approaches the real axis. On the other hand, the upper edges of true quasi-continua quickly converge to their final values when the resolution is increased. We use this observation to define the first criterion for the detection of quasi-continua with \texttt{Legolas}: \textit{a QR-solver locates edges of the quasi-continuum, and with increasing resolution, the upper edge converges whereas the lateral edges are increasingly resolved.}

\begin{figure}[h]
    \plotone{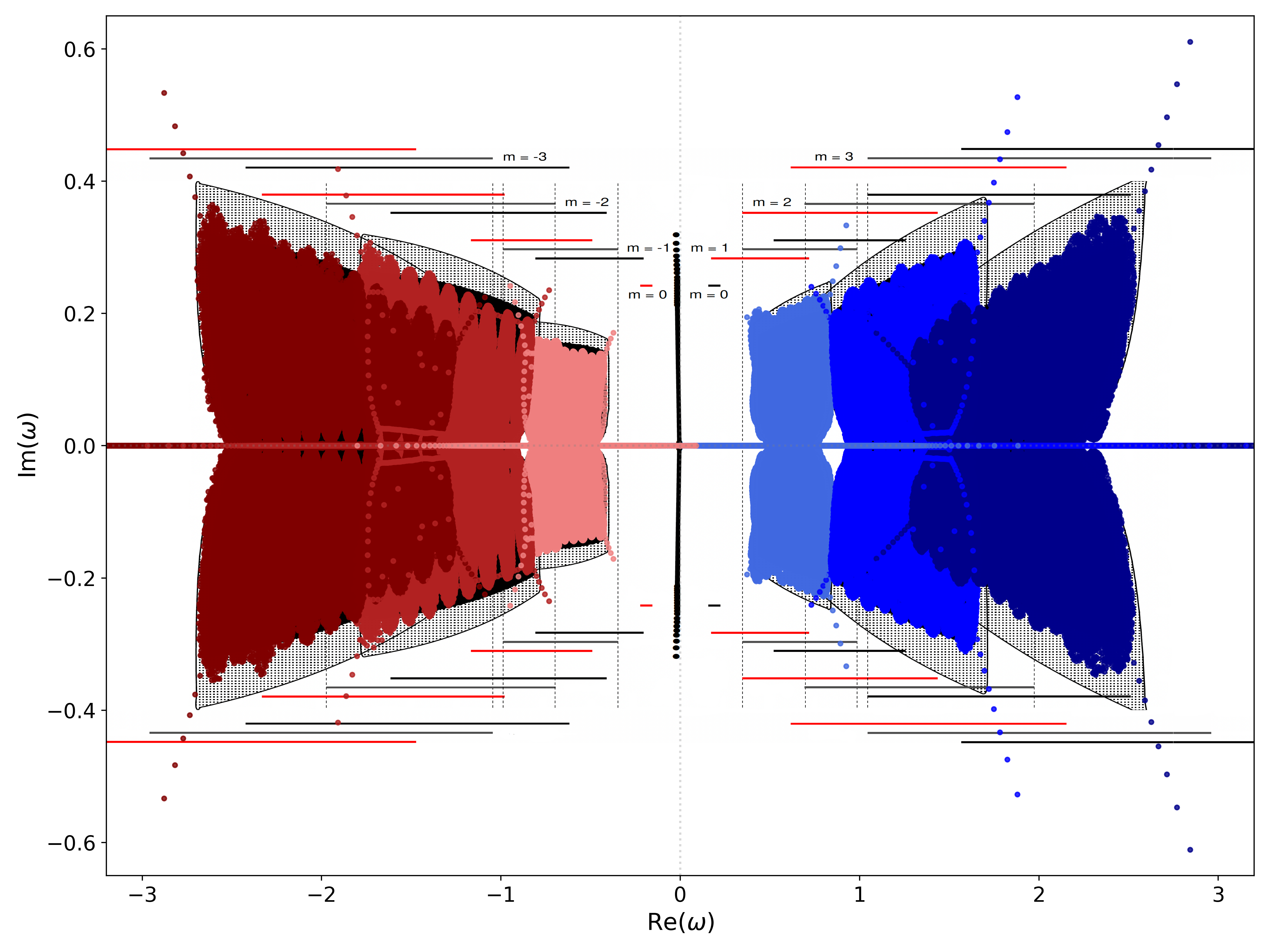}
    \caption{Quasi-continua of SARI modes at various wavenumbers $m\in\{-3,-2,\ldots,3\}$ showing a large collection of unstable frequencies for the same equilibrium. \texttt{Legolas} also calculates the discrete modes clustering to the continua. The ranges of $\Omega_A^\pm$ and $\Omega_0$ are indicated in red/black and gray, respectively. Equilibrium parameters: $\delta = 1$, $\epsilon = 0.045$, $\mu_1=10$, $\beta=10$, and wavenumbers $k/|m| = 50$ (SARIs) and $k=50$ (MRI).
    }
    \label{fig:composite_QC}
\end{figure}

To show that the regions delimited by the QR-invert solver actually contain infinitely many unstable modes, we use the Arnoldi shift-invert solver (\texttt{ARPACK}, \citealt{arpack}), which finds a predefined number of eigenvalues closest to some complex target frequency $\omega_t$. We define a grid of target shift frequencies that sample the 2D region of interest, where we use the quasi-continuum edges found by the QR algorithm. Thanks to the recent upgrades to the \texttt{Legolas} code \citep{legolas2}, this can now easily be done at high radial resolutions: we choose a resolution of $n = 1000$ gridpoints for the equilibria and look for 100 eigenvalues in the neighbourhood of each shift. Figure.~\ref{fig:QC_resolution} (b) shows the result of several of such runs for $m=1$.  An annulus of eigenvalues is found around each shift in the quasi-continuum, with the inner and outer radii depending on the location, resolution, and number of eigenvalues. However, it is clear that the annuli are only artificial: the eigenvalue clouds found by nearby shifts overlap, allowing for the 2D quasi-continua to be filled completely (as we do in Fig.~\ref{fig:composite_QC}). This observation can be used as a second criterion for defining quasi-continua with \texttt{Legolas}: \textit{when applying shift-invert within the bounds determined by the QR-algorithm, arbitrarily many modes can be located in the direct neighbourhood of the shift.}

Panels (c) and (d) of Fig.~\ref{fig:QC_resolution} show three quasi-mode eigenfunctions, sampled at various locations in the quasi-continuum as indicated in Panel (b). All three eigenfunctions $\Re(v_r)$ resemble wave packages having a different number of oscillations and shifted to various locations in the disk. Note how the number of oscillations increases as $\Im(\omega)$ decreases, but the eigenfunctions remain radially localised -- in contrast to the MRI, which becomes more global as the number of oscillations increases. Like the discrete SARI eigenfunctions, each quasi-mode eigenfunction peaks at the local Doppler corotation radius (dashed, gray) and has one or two resonances with the Alfvén continua (red and black, dotted). The eigenmode in Panel (c) has resonances with both continua. At these locations, large skin currents appear that effectively shield the mode from the radial boundaries \citep{GK22}. This eigenmode then represents a truly radially localised wave package moving at the local Doppler speed, insensitive to the boundaries. How local a quasi-continuum mode is, is determined by the separation of these two resonances: if the overlap of $\Omega_A^\pm$ is large, as it is for high-$m$ modes, the separation is very small and the modes are hence radially very localised (but still global in nature). As for the two quasi-modes in Panel (d), the orange and green eigenfunctions only resonate with the forward (red) and backward (black) Alfvén continua, respectively. In this sense, modes like these are the quasi-continuum equivalent of discrete SARI modes, where the second Alfvén resonance can now be either inside the domain or not. An inspection of the other eigenfunctions (not shown) reveals that the Lagrangian perturbations along and perpendicular to the magnetic field are the largest, highlighting the slow/Alfvénic nature of the instability.

The huge potential of quasi-mode SARIs to govern instability in accretion disks becomes clear when plotting composite spectra for several wavenumbers, where the Doppler range moves for every $m$. In Fig.~\ref{fig:composite_QC}, we show spectra for modes with wavenumbers $m\in\{-3,-2,\ldots,3\}$ and wavenumbers $k/|m| = 50$ for the SARIs and $k=50$ for the MRI, recreating Fig.~20 of \citet{GK22} using \texttt{Legolas}. We use the combined efforts of the QR algorithm to find the quasi-continuum edges and the Arnoldi shift-invert algorithm to fill the 2D regions. The colour code adopted here will be used in the remainder of this paper: the $m=0$ MRI in black and the $m<0$ counter- and $m>0$ co-rotating SARIs in red and blue, respectively, to give the Doppler shift its usual connotation. We find an excellent correspondence with the $W_\text{com} \leq 10^{-12}$ regions delimited by \texttt{ROC}. The Alfvén continua (red and black) and Doppler range (gray) are indicated for each value of $m$, and it is clear that each quasi-continuum is limited to the Doppler range. The furthest edges of the $m=\pm3$ continua fall out of this frame at around $\sigma = -3.5$ and $3.8$, respectively. With \texttt{Legolas}, we additionally obtain the discrete SARI modes clustering towards the continua, as well as stable Alfvénic modes (visible on the real axis for $m=\pm1$). Note again the up-down symmetry of the spectrum, a feature of ideal MHD \citep{GKP19}. This figure underlines that a typical equilibrium has a huge range of unstable perturbations over many wavenumbers, with perturbations that can be radially localised anywhere in the disk! If a linear MHD view suffices to explain the cause of turbulence, this abundance of entire 2D regions of near-eigenmode SARIs that grow exponentially and experience differential rotation in the disk is a much more convincing argument than the existence of a few isolated unstable MRIs.

\section{SARI and MRI in accretion disk equilibria} \label{sec:param_study}

The previous Section made it clear that allowing for modes with $m\neq0$ has important implications for accretion disk stability. We independently confirmed the existence of continuous regions of quasi-modes using \texttt{Legolas}, and pointed out how the axisymmetric MRI differs from its non-axisymmetric counterparts. In this Section, we set out to systematically explore the stability of inner/outer SARIs and MRI in various disk equilibria, thereby comparing the fastest-growing modes over a range of wavenumbers $m$. In Sect.~\ref{sec:vertical_localisation}, we first look at the influence of the vertical wavenumber on MRI/SARI growth. We then explore the influence of the magnetic field orientation (Sect.~\ref{sec:field_orientation}), and field strength and plasma-beta (Sect.~\ref{sec:magnetisation}). These results are then compared to predictions from analytical theory in Sect.~\ref{sec:instability_criterion}. In Sect.~\ref{sec:param_spectra}, we use this knowledge to explicitly obtain a quasi-continuum of extremely localised modes.

Before exploring the instabilities further numerically, it is useful to state the MRI/SARI instability criterion and analytical maximum growth rate predictions derived by \citet{GK22}. This upper bound for growth, obtained from the global problem (their Eq.~(84)), improves on an upper bound derived under the WKB approximation \citep{BH92b}:
\begin{equation} \label{eq:numax_GK22}
    \nu_\text{max}(r) = \sqrt{\frac{1}{2}\sqrt{\kappa_\text{e}^4 + 16\Omega^2\omega_A^2} - \frac{1}{2}\kappa_\text{e}^2 - \omega_A^2} \lesssim \sqrt{3} |\omega_A|,
\end{equation}
which yields the well-known instability criterion \citep{BH91,ogilvie_pringle,BH98,blokland05},
\begin{equation} \label{eq:instability_criterion}
    0 < \omega_A^2 < 4\Omega^2 - \kappa_\text{e}^2 = 3\Omega^2, \qquad \omega_{A \text{cut-off}} = \sqrt{3} \Omega.
\end{equation}
Here, the squared epicyclic frequency $\kappa_e^2 = 4\Omega^2 + r\left(\Omega^2\right)' = \Omega^2$ for our Keplerian rotation profile. Note that the Alfvén frequency $\omega_A = \left(\frac{m}{r} B_\theta + k B_z\right) / \sqrt{\rho}$ contains both wavenumbers $k$ and $m$, and it can now go through zero and become negative for $m<0$, in contrast to the situation for the MRI. If $\omega_A$ vanishes and the lower bound is violated, the modes should stabilise. The same should happen if the upper bound is violated because the field gets too strong or the wavenumbers too large -- physically, magnetic tension then stabilises modes with wavelengths that are sufficiently small. Equation~\eqref{eq:numax_GK22} can also be differentiated with respect to $\omega_A^2$ to obtain the maximum growth rate over all wavenumbers at a fixed radius:
\begin{equation} \label{eq:numax_wavenumbers}
    \nu_\text{max}(r) = \frac{3}{4} \Omega, \qquad \text{for}\quad\omega_{A\text{max}} = \sqrt{\frac{15}{16}} \Omega,
\end{equation}
which exactly matches the well-known local MRI prediction \citep{BH98} but for a completely general field orientation. We will compare the above predictions to our numerical results in Sect.~\ref{sec:instability_criterion}.

The dimensionless parameters $\epsilon$, $\beta$, $\mu_1$ and $\delta$ defined in Sect.~\ref{sec:model_equilibrium} completely determine an ideal MHD equilibrium with cylindrical flow and gravity, representing an accretion disk. \citet{GK22} take $v_{A1} = B_1 = \epsilon \ll 1$ and $p_1 \sim \epsilon^2\beta \ll 1$, and $k^2r^2 \gg m^2$ so that the equilibrium state represents a near-Keplerian ($\Omega_1 \approx 1$), weakly magnetised, thin accretion disk (vertical scale height $H \sim c_S = \frac{\gamma}{2} v_{A1}^2 \beta \ll 1$ \citep{pringle1981}). Super-Alfvénic rotation then follows, since $v_{A1} \ll 1 \sim m\Omega_1$. These assumptions allow for an analytical treatment of the global eigenvalue problem, but a numerical code like \texttt{Legolas} is not limited to those regimes. Nevertheless, we will find that in most cases, the analytical growth rate predictions are still sufficient to understand the stability of the modes (Sect.~\ref{sec:instability_criterion}).

The values of the parameters are fixed to one of the two following sets throughout this Section: 
\begin{itemize}
    \item \textbf{Case (a):} $\epsilon = 0.0141$, $\mu_1 = 1$, $\beta = 100$, $\delta = 1$, and $k = 50$, corresponding to a weak field and pitch angle $\varphi = 45^\circ$ (see Sect.~\ref{sec:model_equilibrium}); or
    \item \textbf{Case (b):} $\epsilon = 0.045$, $\mu_1 = 10$, $\beta = 10$, $\delta = 1$, and $k = 50$, corresponding to a slightly stronger, predominantly toroidal field ($\varphi \approx 85^\circ$), 
\end{itemize}
where one or more parameters are varied. Both sets of parameters have a sizeable $B_\theta$ component -- indeed, there is a growing set of evidence that significant toroidal fields may be generated in accretion disks \citep{das18, begelman22}. The main differences between the two cases are then an equipartition between the poloidal and toroidal fields versus dominant poloidal field, and a weak field versus a slightly less weak field. In both cases, $\Omega_1 \approx 0.987$. 

\texttt{Legolas} works in dimensionless units, which are fixed in all runs by the following default choices: $B_u = 10$ G for the magnetic field, $L_u = 10^9$ cm as the length unit, and $T_u = 10^6$ K for temperature. The spectra of Figs.~\ref{fig:MRI_SARI}, \ref{fig:QC_resolution}, \ref{fig:composite_QC} and \ref{fig:QC_m-10} are given in these dimensionless values. The growth rates in Figs.~\ref{fig:parameter_k}--\ref{fig:parameter_beta} are additionally normalised by the inner disk rotation frequency $\Omega_1$. All spectra in those Figures are calculated at a resolution of $200$ base gridpoints.

\subsection{Vertical localisation} \label{sec:vertical_localisation}
It is well known from local linear theory that the MRI has a cut-off value for the vertical wavenumber $k$ \citep{BH91,BH98} - the strength of the poloidal field component determines the smallest vertical wavelength not stabilised by magnetic tension. We first investigate if this generally applies to SARIs in a global eigenfunction description, and at which vertical wavelengths the growth is maximal by quantifying the most unstable modes from numerical spectra obtained with \texttt{Legolas}.

\begin{figure}
    \centering
    \includegraphics[width=0.6\linewidth]{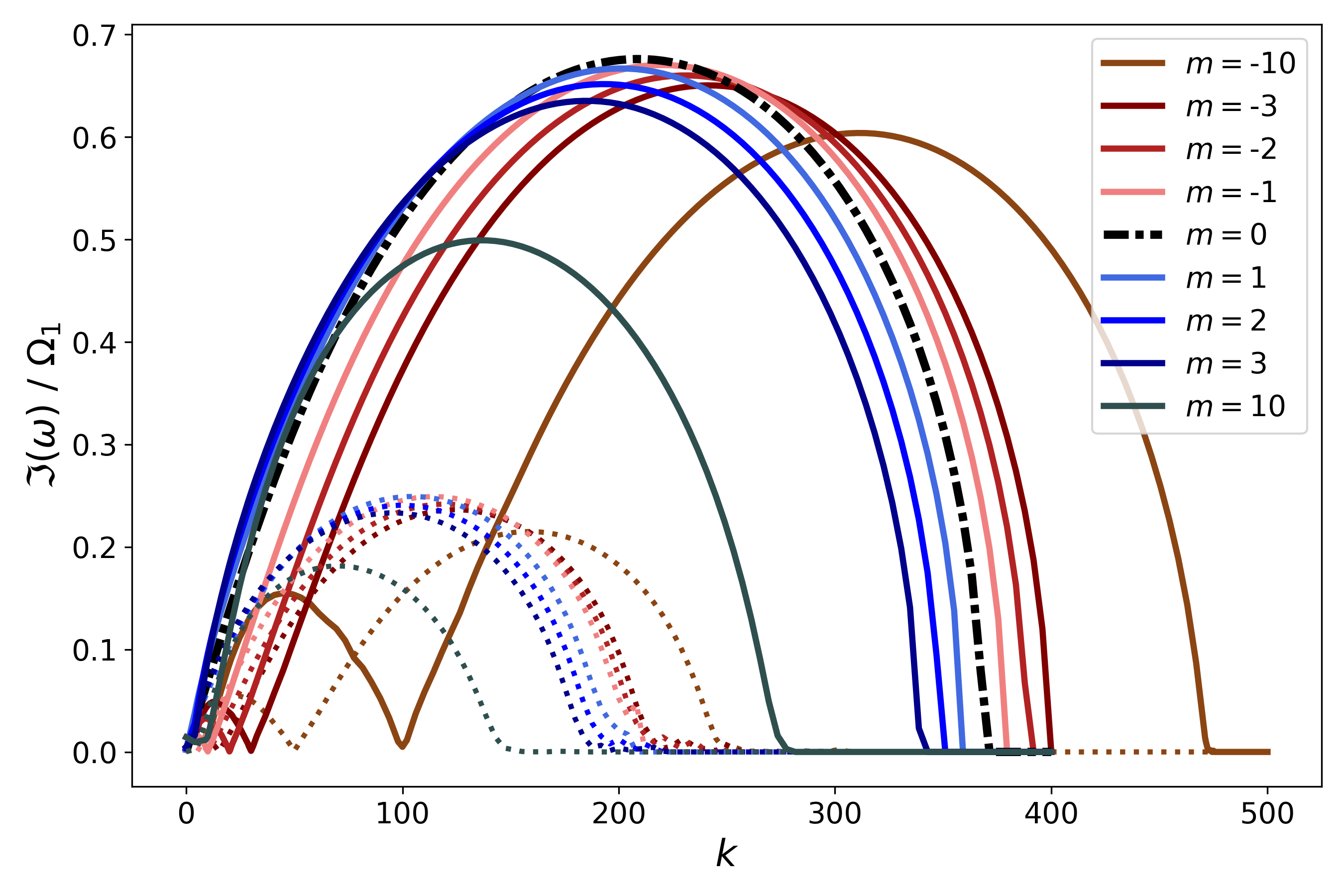}
    \caption{Normalised growth rates for MRI (black, dashdot) and SARI (inner: solid, outer: dashed) over a range of vertical wavenumbers $k$ in Case (b) (dominant azimuthal field). Both $k_\text{cut-off}$ and $k_\text{max}$ increase as $m$ decreases. The outer growth rates are scaled-down versions of the inner curves. Cusps in the growth rate curves, e.g. at $k=100$ for the inner $m=-10$ SARI, signify local stabilisation if $\omega_A$ vanishes.
    }
    \label{fig:parameter_k}
\end{figure}

Figure~\ref{fig:parameter_k} shows how the growth rates vary with $k$ for Case (b) ($\varphi=85^\circ$), which is more illustrative here than Case (a). The growth rates of the $m=0$ MRI are shown in black, and those of the $m>0$ and $m<0$ SARIs in blue and red, respectively, with the inner SARIs in solid lines and the outer branch dotted. One immediate observation is that all instabilities have a clear cut-off wavelength and reach maximum growth for intermediate wavelengths. Both the cut-off wavenumber $k_\text{cut-off}$ and the wavenumber of maximal growth $k_\text{max}$ are ordered by $m$ and significantly increased for counter-rotating $m<0$ SARIs, which allow for much smaller unstable vertical wavelengths, as was also found by \citet{begelman22}. Maximum growth rates are comparable for the MRI and inner SARI, but the outer SARIs consistently have lower growth rates. The latter growth rate curves largely follow those of the MRI and inner SARIs, but scaled down by a factor of 2 horizontally and a factor of around 3 vertically. The reason for this scaling will be explained in \ref{sec:instability_criterion}. Strikingly, for $m<0$ SARIs, there is an intermediate value of $k$ for which the growth rates vanish before the maximum is reached. This is most prominent for $m=-10$, where the inner branch is stabilised at $k=100$ and the outer branch at $k=50$. At these wavenumbers, the Alfvén frequency vanishes locally, i.e. $\omega_A(r_*)=0$, at the corotation location $r_*$, which violates the SARI instability criterion given by Eq.~\eqref{eq:instability_criterion}. This highlights that the magnetic field is essential for instability, and instability does not prevail for those wavenumbers where the wave is directed perpendicularly to the field ($\mathbf{k} \cdot \mathbf{B} = 0$). It is important to bear in mind that the range of allowed $k$ is limited in a realistic, 2D disk with a vertical extent and hence a quantisation on $k$ or at least an upper limit for vertical wavelengths. Hence, the range of $k$ where the modes are physical is determined by $H \sim c_S$, which gives in this case $k \gtrsim 30$. Case (a) (not shown) has much smaller values of $k_\text{max}$ and $k_\text{cut-off}$ because of the stronger vertical field component. The variation with $m$ is similar to Case (b) but less pronounced: because the field has a larger vertical component ($B_{z1}=B_{\theta1}$), the influence of the azimuthal wavenumber $m$ is reduced and one needs to go to more negative $m$ to significantly increase $k_\text{cut-off}$. 

\subsection{Field orientation} \label{sec:field_orientation}
Most analytical treatments of accretion disk instabilities consider the limiting cases of a dominant poloidal background field \citep{BH91,BH98, mamatsashvili13} or a dominant toroidal background field \citep{terquem_papaloizou,ogilvie_pringle,das18}. However, a global spectroscopic approach can treat any completely general equilibrium with radial variation \citep{keppens2002,blokland05}, where in our case the orientation and strength of the magnetic field can be smoothly varied by the parameters $\mu_1$ and $\epsilon$. In what follows, we consider inverse pitch angles $\varphi := \arctan{B_{\theta1}/B_{z1}}$ between $0^\circ$ and $90^\circ$, where the former corresponds to a purely poloidal (vertical) field and the latter to a purely toroidal (azimuthal) field, and calculate SARI and MRI growth rates at a fixed vertical wavenumber $k=50$.

\begin{figure}
    \centering
    \includegraphics[width=0.49\linewidth]{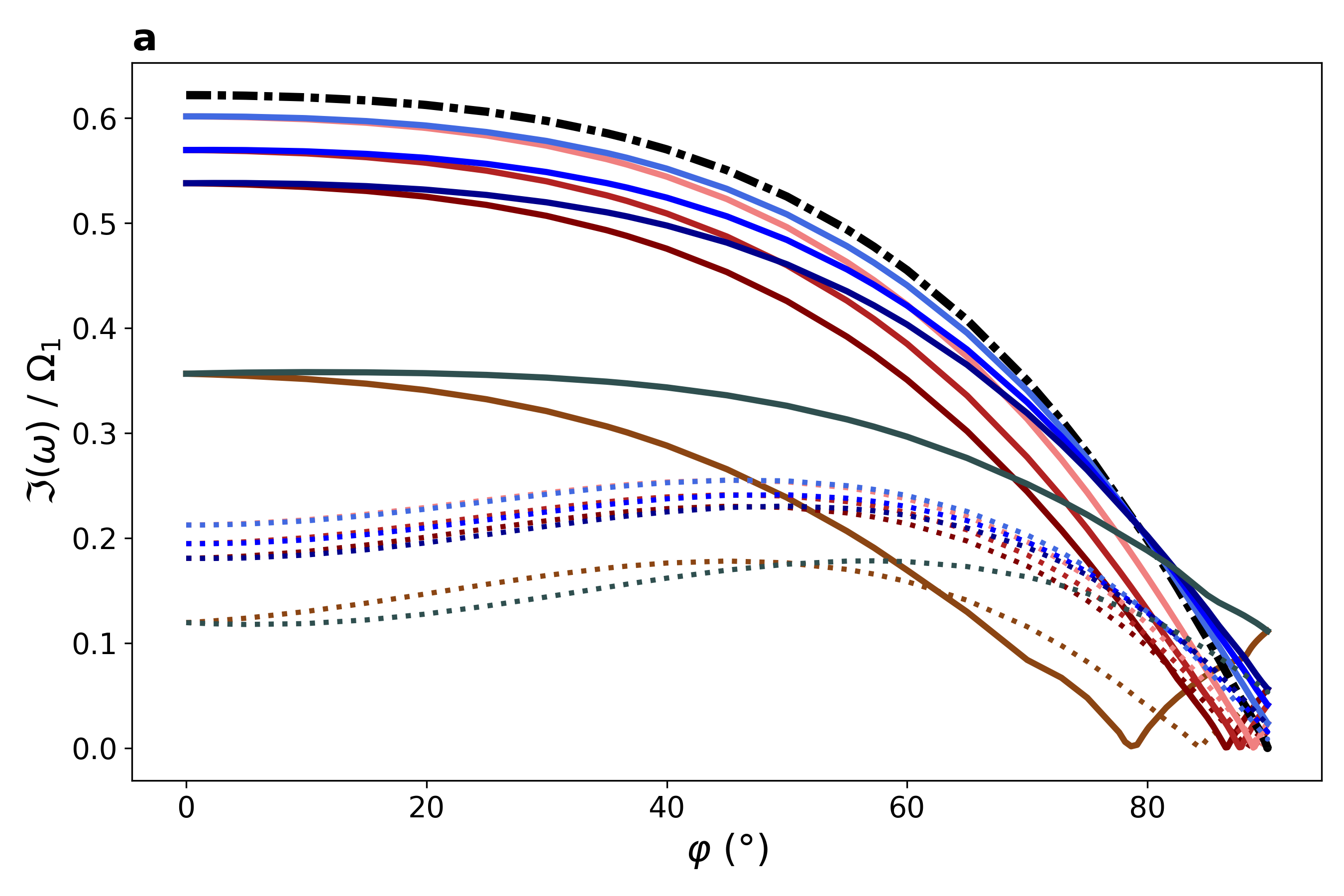}
    \hspace*{0.1 cm}
    \includegraphics[width=0.49\linewidth]{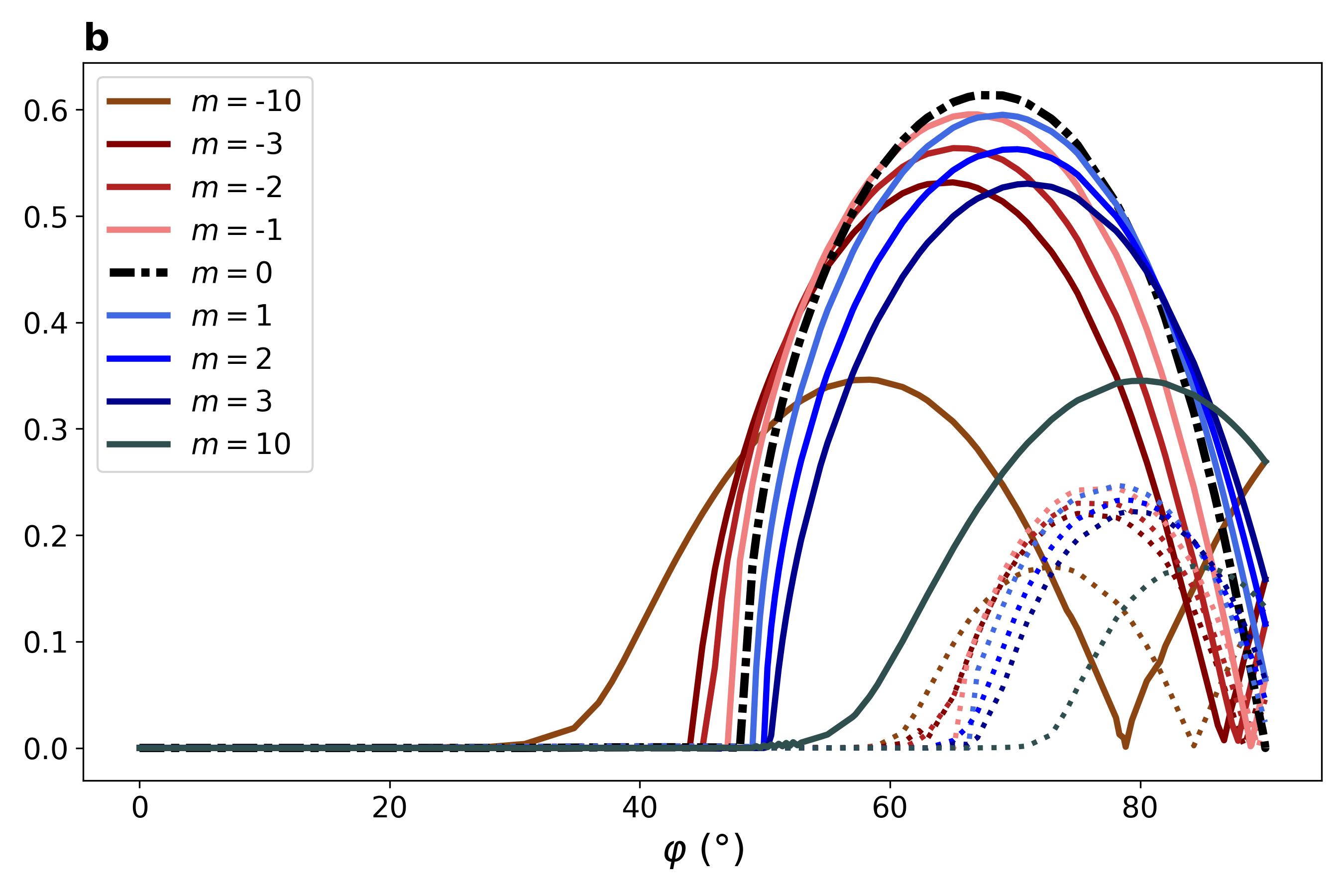}
    \caption{Normalised growth rates for MRI (black, dashdot) and SARI (inner: solid, outer: dashed) for purely vertical ($\varphi=0^\circ$) to purely azimuthal ($\varphi=90^\circ$) background field for Case (a) and Case (b) at $k=50$. Note how a slight increase in field strength in Panel (b) greatly reduces the allowed poloidal component. MRI stabilises in a purely toroidal field; SARIs become more unstable.}
    \label{fig:parameter_mu}
\end{figure}

Figure~\ref{fig:parameter_mu} (a) shows the growth rates for Case (a) (weak field and $\beta = 100$). The MRI and low-$m$ SARI are the fastest-growing modes whenever there is a significant $B_z$-component present, with inner and outer SARI growth rates decreasing with $|m|$. All growth rates drop significantly as the field becomes more toroidal, but the SARI growth rates remain finite whereas the MRI is stabilised completely as $\varphi\rightarrow 90^\circ$: with $B_z \rightarrow 0$ and $m=0$ there is no way for the instability to couple to the toroidal component, and $\omega_A = 0$ everywhere \citep{BH92b}. For a purely toroidal field, the growth rates increase with $|m|$. This is understood from Fig.~\ref{fig:parameter_k}: an equivalent figure for fixed $k$ and varying $m$ would show an equivalent growth rate curve for $m$ with an $m_\text{max}$ and $m_\text{cut-off}$. Hence, for values of $|m|$ below $m_\text{max}$, the growth rates increase with $|m|$. As in Fig.~\ref{fig:parameter_k}, for every $m<0$ there is a value of $\varphi$ at which $\omega_A(r_*) = 0$ near the inner or outer boundary and as a result the respective inner or outer mode that resonates with this frequency is locally stabilised.

Figure~\ref{fig:parameter_mu} (b) of Case (b) (slightly stronger field and $\beta=10$) shows quite a different situation: at $k=50$, both MRI and SARI are suppressed when the vertical field component dominates, and they are only activated for a sizeable toroidal component. The `cut-off' inverse pitch angle is seen to decrease as $m$ becomes increasingly negative: the relative strength of the poloidal field allowed for instability increases. Comparing with Case (a), we conclude that for $\epsilon=0.014$, the vertical component is able to support instability at $k=50$ instability, but $\epsilon=0.045$ is too strong unless $m$ is very negative. The cut-off being less sharp for $m=\pm 10$ SARIs is a resolution effect. For a purely toroidal field, the MRI is again suppressed and SARIs reach comparably large growth rates, again increasing with $|m|$ in this range of azimuthal wavenumbers. Again, $m<0$ SARIs can be stabilised locally. The maximum growth rates of all instabilities are comparable to those of Case (a), but the stronger azimuthal field increases the SARI growth rates in the limit $\varphi \rightarrow 90^\circ$.

\begin{figure}
    \centering
    \includegraphics[width=0.6\linewidth]{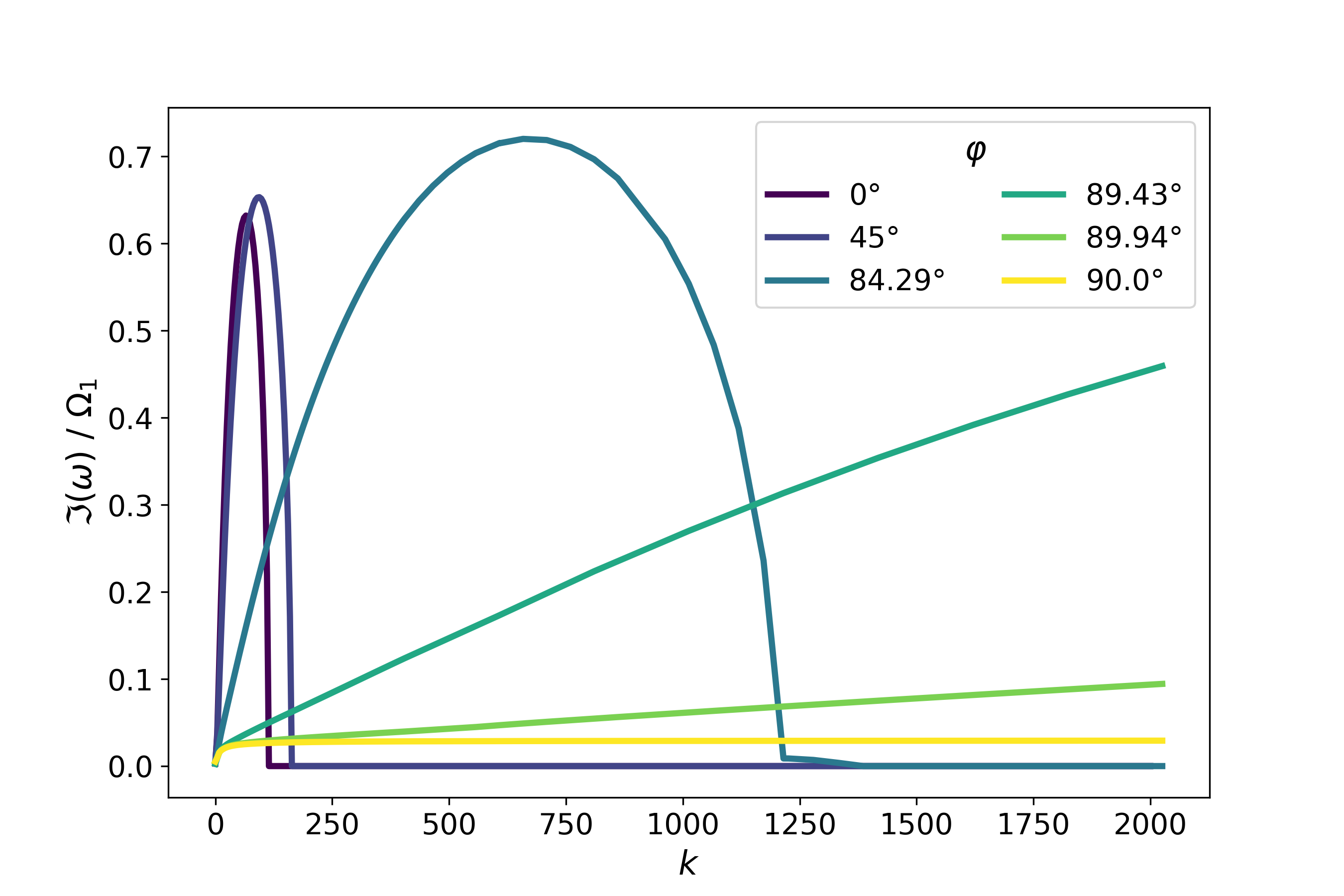}
    \caption{Growth rates of the $m=1$ inner SARI peak and cut-off at increasingly large $k$ when the field becomes nearly toroidal but this extreme localisation breaks down for even a small poloidal component. However, growth rates remain finite for small $k$ in the limit $\varphi \rightarrow 90^\circ$.}
    \label{fig:parameter_mu_and_k}
\end{figure}

\citet{ogilvie_pringle} performed a global analysis of a cylindrical accretion disk model with a purely toroidal background field. They found that instability prevails for arbitrary values of the wavenumbers $k$ and $m$ and moreover, that fastest growth happens on the smallest scales, implying that there is no cut-off vertical wavelength $k$. \citet{terquem_papaloizou} also came to this conclusion when finding modes growing on increasingly localised scales, eventually limited only by dissipative scales. Figure~\ref{fig:parameter_mu_and_k} features the growth rate variation with $k$ for six field orientations and $m=1$ fixed. If the poloidal component dominates ($\varphi \leq 45^\circ$), the $k_\text{cut-off}$ varies only slightly below $k=200$. As the relative strength of the toroidal component increases, $k_\text{cut-off}$ moves further away to values above $k=1000$, and eventually to arbitrarily large values as $\varphi \rightarrow 90^\circ$. A similar observation can be made for $k_\text{max}$. This confirms that indeed, for all practical purposes the vertical scale of the fastest-growing modes becomes arbitrarily small for a toroidal field. However, this extreme localisation immediately breaks down when the field gets a small poloidal component, as was also noted by \citet{BH92b}. Additionally, the growth rates remain finite  in the limit of a purely toroidal field even for low $k$ because of the non-axisymmetric coupling to the toroidal component. For an axisymmetric mode like the MRI, this is not the case. 

\subsection{Magnetisation} \label{sec:magnetisation}
Two of the essential assumptions for SARI modes (and for the standard MRI in \citet{BH91}) are a weak field, i.e. $v_{A1} \ll 1$ and a thin disk with scale height $H \sim c_{S1} = \frac{\gamma}{2} v_{A1}^2 \beta \ll 1$, so that near-Keplerian disks rotate super-Alfvénically and the background is close to incompressible. We now investigate fields from weak to dynamically relevant, modeling disks with near-Keplerian to strongly sub-Keplerian rotation speeds. 

\begin{figure}
    \centering
    \includegraphics[width=0.49\linewidth]{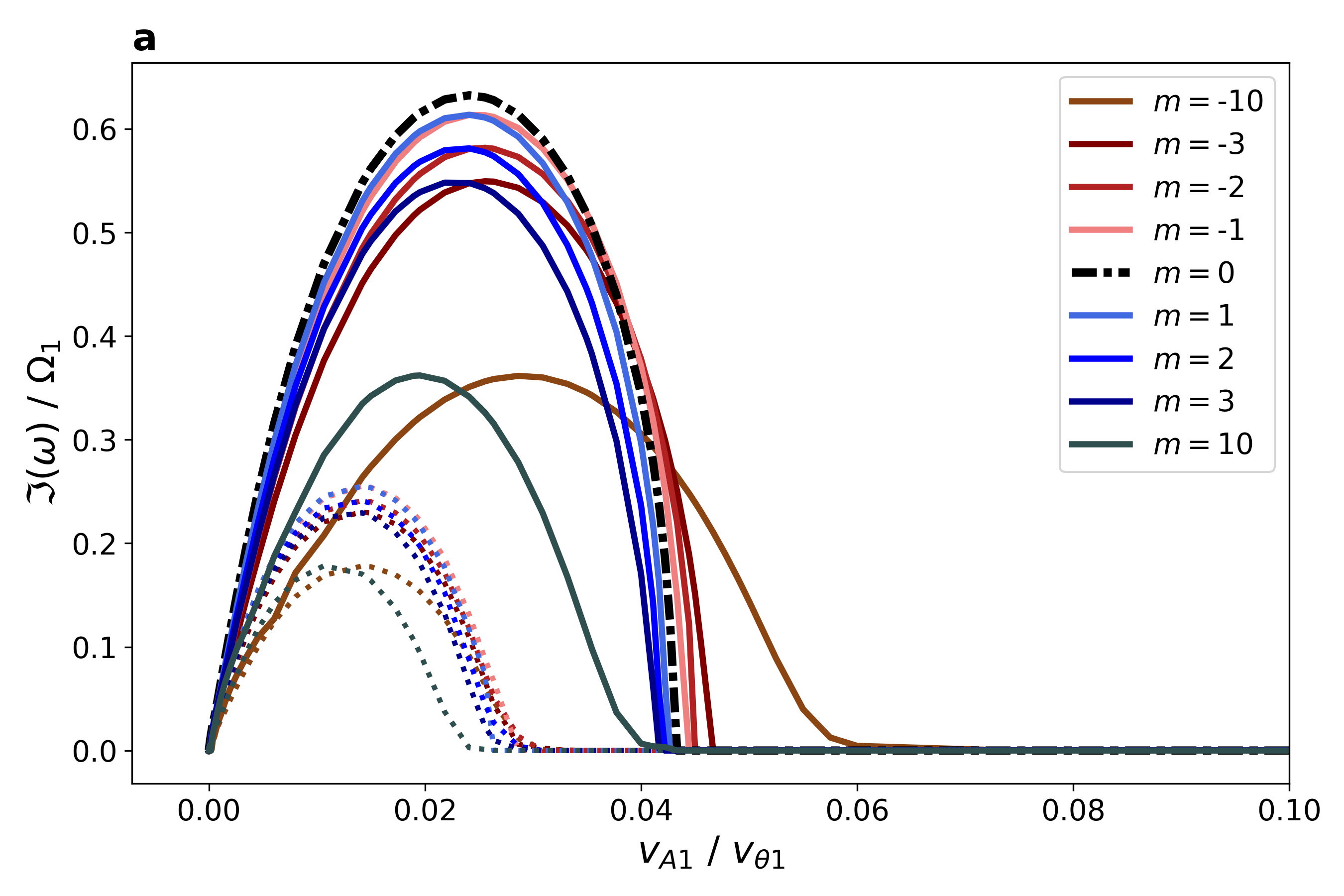}
    \hspace*{0.1 cm}
    \includegraphics[width=0.49\linewidth]{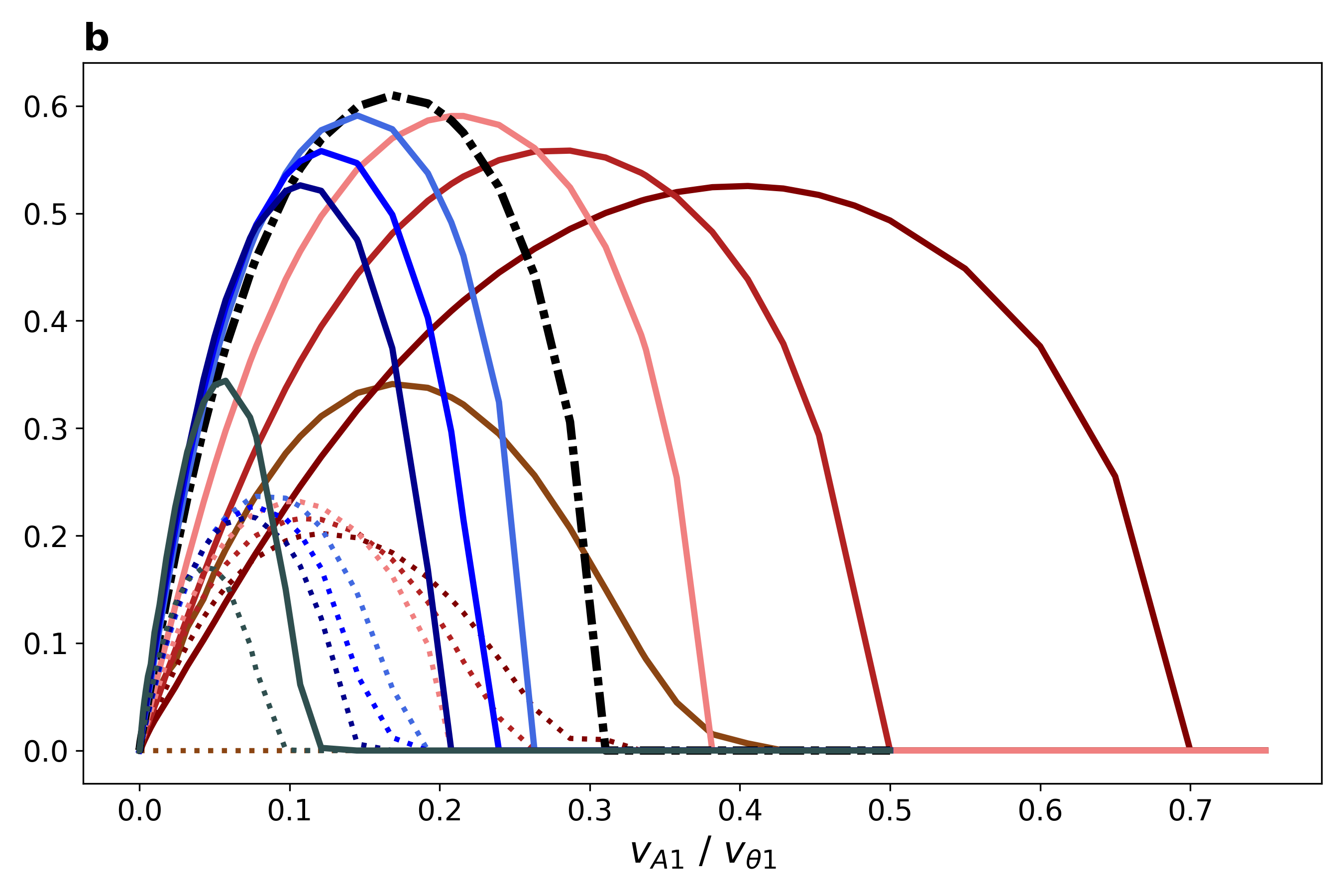}
    \caption{Normalised growth rates of MRI (black, dashdot) and SARI (inner: solid, outer: dashed) as a function of normalised inner Alfvén velocity ($v_{A1} = B_1$). A helical field (a) provides only a small range for instability at $k=50$, in contrast to a dominant toroidal field (b), where especially the $m<0$ SARIs are maximally unstable at much larger field strengths.}
    \label{fig:parameter_epsilon}
\end{figure}

Figure~\ref{fig:parameter_epsilon} shows the MRI and SARI growth rates as a function of the Alfvén speed $v_{A1} = \epsilon$ normalised by the inner rotation speed $v_{\theta1} = r_1\Omega_1$, with the most unstable inner and outer SARIs again denoted by solid and dashed lines, respectively. Panel (a) and (b) again show Case (a) with $\varphi = 45^\circ$ and Case (b) with a dominant toroidal field, respectively, both with $k=50$ and $\beta > 10$. An immediate result is that for every $k,m$ there is a bounded range of field strengths at which instability operates. For Case (a), all growth rates are maximal for $v_{A1} \approx 0.02 v_{\theta1}$, well in the extremely super-Alfvénic regime, but for Case (b) the range of $v_{A1}$ in which instability prevails for the different $m$ is much larger, with $v_{A1} / v_{\theta1}$ close to unity for $m<0$. The $m=-3$ SARI almost doubles the range of instability as compared to the MRI at this value of $k$. In both panels, the values of $v_{A1\text{max}}$ and $v_{A1\text{cut-off}}$ shift with the value of $m$ in a distinctive way. In Panel (a), both $v_{A1\text{max}}$ and $v_{A1\text{cut-off}}$ decrease as the wavenumbers go from $m<0$ to $m>0$. In Panel (b), the shifts are more pronounced and do not obey the same ordering: remarkably, the $m=-10$ SARI is cut off at significantly weaker fields than other $m<0$ SARIs, and its $v_{A1\text{max}}$ corresponds roughly with that of the MRI. This pattern will be explained from the SARI instability criterion in Sect.~\ref{sec:instability_criterion}. Note that in both cases the outer SARIs again behave very similarly to the inner SARIs, but with lower growth rates and stricter cut-off. In Case (b), the outer $m=-10$ SARI is stabilised, as $k=50$ is the vertical wavenumber where $\omega_A(r_2) = 0$ (this is also clear from Fig.~\ref{fig:parameter_k}). Note that the normalisation factor $\Omega_1$ is itself dependent on $\epsilon$ (Sect.~\ref{sec:model_equilibrium}). This parameter can become much smaller than its Keplerian value $\Omega_1=1$ for the strongest fields in Case (b), going down to $\Omega_1 < 0.3$. For such sub-Keplerian disks, the inner disk is magnetically supported instead of rotationally.

\begin{figure}
    \centering
    \includegraphics[width=0.49\linewidth]{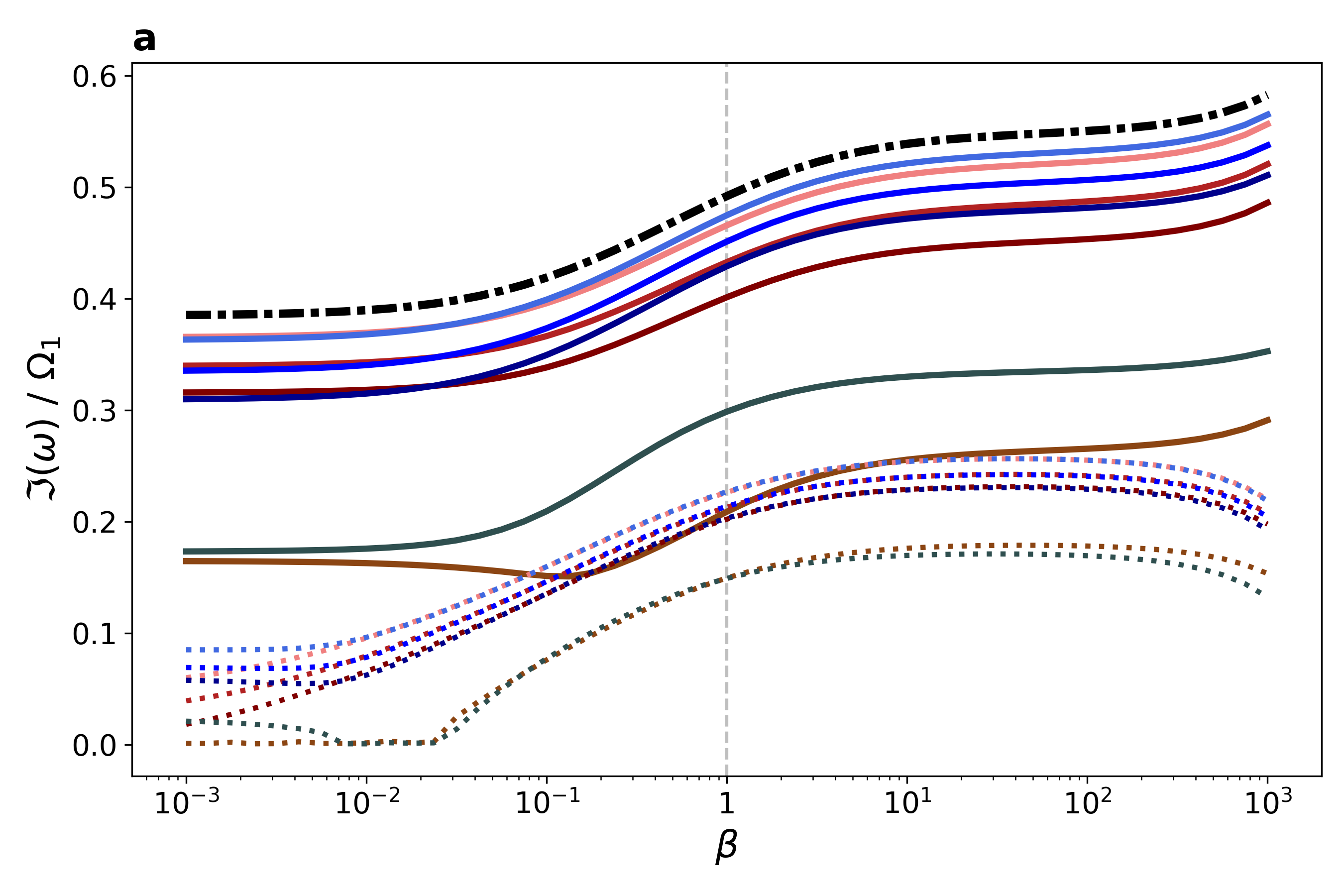}
    \hspace*{0.1 cm}
    \includegraphics[width=0.49\linewidth]{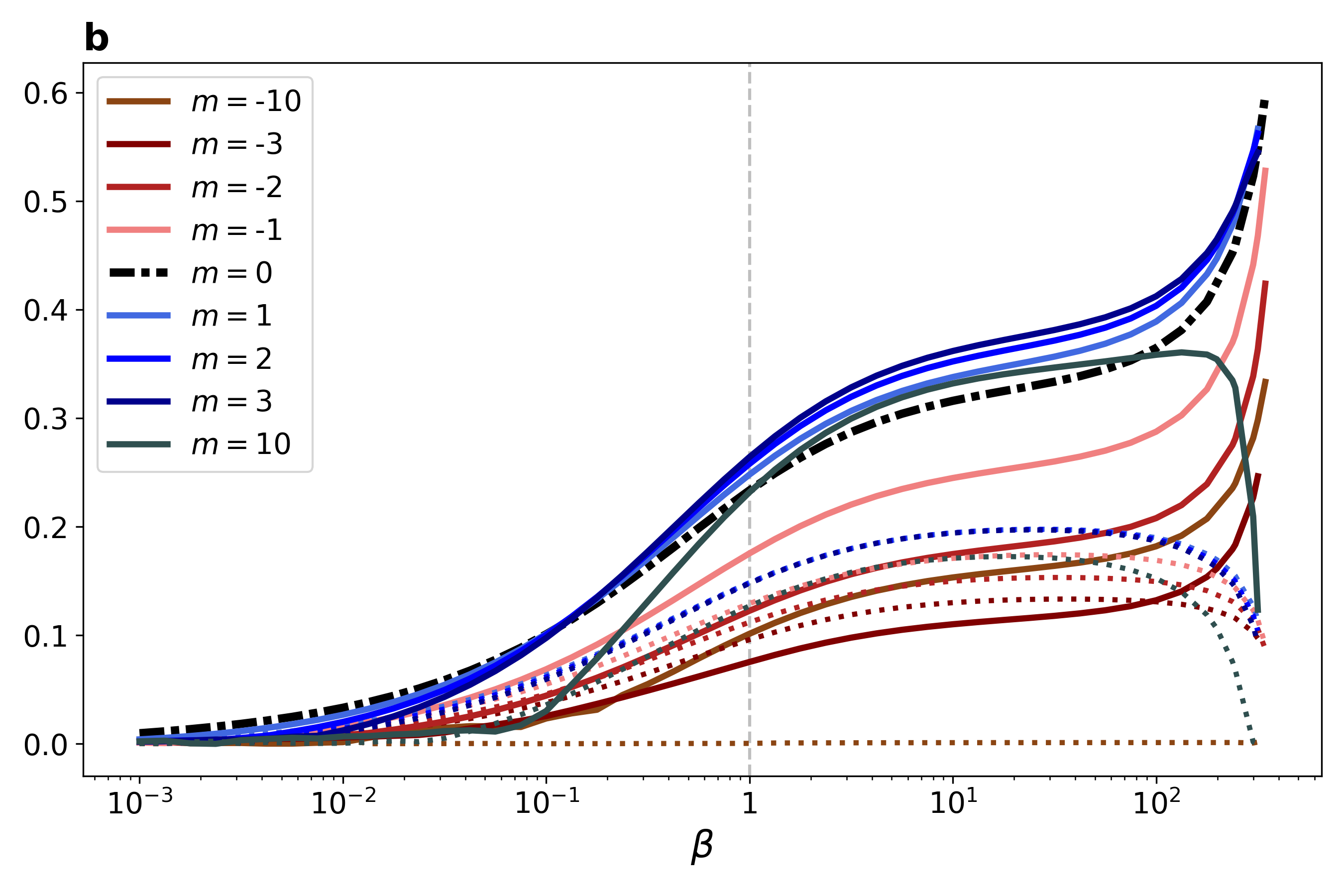}
    \caption{Normalised growth rates of MRI (black, dashdot) and SARI (inner: solid, outer: dashed) varying with plasma-beta at $k=50$. In suprathermal fields, growth remains finite for a helical field (a) but converges to $0$ for a dominant toroidal field (b). In Panel (b), the MRI is not the most unstable mode at this value of $k$.}
    \label{fig:parameter_beta}
\end{figure}

The field strength can also be considered relative to the gas pressure through varying the plasma-beta. Although the standard MRI was first obtained for disks threaded by a weak, subthermal field \citep{BH91}, it has been shown to prevail in suprathermal disks as well \citep{das18}, motivated by simulations where a strong suprathermal toroidal field is generated ($\beta \sim 0.1$ to $1$) \citep{bai_stone13}. Figure~\ref{fig:parameter_beta} (a) shows how the MRI and SARI growth rates depend on the plasma-beta for our Case (a) with equal poloidal and toroidal field components. The MRI is the most unstable mode, closely followed by the inner SARI, with growth rates decreasing with $|m|$. $\pm m$ SARI growth rates almost coincide for $\beta < 1$ but are separated for subthermal fields, which becomes more apparent as $|m|$ increases. For these equilibria, suprathermal magnetic fields ($\beta > 1$) produce slightly higher growth rates than subthermal fields ($\beta < 1$). The outer SARIs show a similar trend, but the $\pm m$ growth rates almost coincide for subthermal fields. They diverge, however, in the limit $\beta \rightarrow 0$, where the co-rotating modes retain a finite growth rate but the counter-rotating modes have vanishing growth rate. Note that the $m=-10$ outer mode is stabilised completely for $\beta \rightarrow 0$, while the $m=10$ outer mode is briefly stabilised and then reappears. However, a closer inspection of the spectra for these equilibria (not shown) reveals that these modes are not SARIs: the slow continua are separated from the Alfvén continua and have their proper overlap, and the outer eigenfunctions are not confined to the region between the resonances with the Alfvén continuum. They could be related to the hybrid Alfvén/slow modes described in \citet{das18}. Figure~\ref{fig:parameter_beta} (b) shows quite a different behaviour for Case (b) with a slightly stronger dominant toroidal field. All modes show a large variation in growth rates and the MRI is not the most unstable mode over most of the $\beta$ range. All growth rates become low for $\beta \rightarrow 0$. Note the sudden drop in growth of the $m=10$ modes at large values of $\beta$, while other growth rates drastically increase. This is a result of the normalisation, where $\Omega_1$ depends critically on $\beta$. When the disk becomes strongly sub-Keplerian, $\Omega_1 \ll 1$. It is expected that for values of $\beta$ where $\Omega_1 \rightarrow 0$, the growth rates will remain finite and suddenly vanish like those for the $m=10$ modes. In contrast to most of our results shown above, growth rates of the outer $m=-3$ SARI are slightly higher than those of the inner SARI over almost the entire parameter range. Indeed, the wavenumber $k=50$ is close to that where the $m=-3$ inner mode is locally stabilised (Fig.~\ref{fig:parameter_k}), and hence it has a reduced growth rate. The $m=-10$ outer mode is again fully stabilised at this wavenumber.

\subsection{Analytical predictions} \label{sec:instability_criterion}

We now revert back to the MRI/SARI instability criterion and maximum growth rate to explain the results of the previous Sections and test the analytical predictions.

\subsubsection{Instability criterion}
Much of the SARI behaviour can be explained by the instability criterion \eqref{eq:instability_criterion}. Contrary to the $m=0$ MRI, the toroidal field plays a significant role in the instability of non-axisymmetric modes. An important consequence is that the perturbations can couple to the toroidal field: for any given value of $k$, there exists a negative $m$ that makes $\omega_A$ low enough to satisfy the criterion. A second consequence is that for certain wavenumber combinations, $\omega_A(r_*)=0$ locally, hence violating the lower end of the instability criterion. Both of these possibilities are completely absent for the MRI. 

A direct consequence of the first observation is a shift in cut-off vertical wavenumber with $m$ as found in Fig.~\ref{fig:parameter_k}. From Eq.~\eqref{eq:instability_criterion}, it is clear that the cut-off vertical wavenumber shifts to larger values compared to the axisymmetric MRI for $m < 0$ and to smaller values for $m > 0$. For a sizeable poloidal field, this effect is small: for Case (a), which is not shown, all cut-offs are in the interval $[150,175]$ for $|m| \leq 10$. For a dominant poloidal field, the effect is more noticeable: according to the analytical predictions, $k_\text{max}$ and $k_\text{cut-off}$ shift to large values when $m$ becomes arbitrarily negative. Furthermore, the wavenumber $k$ where $\omega_A$ vanishes locally shifts to larger values as $m$ becomes increasingly negative. For increasingly positive $m$, the range of allowed $k$ decreases due to stabilisation by the increasing magnetic tension. 

\begin{figure}
    \centering
    \includegraphics[width=0.6\linewidth]{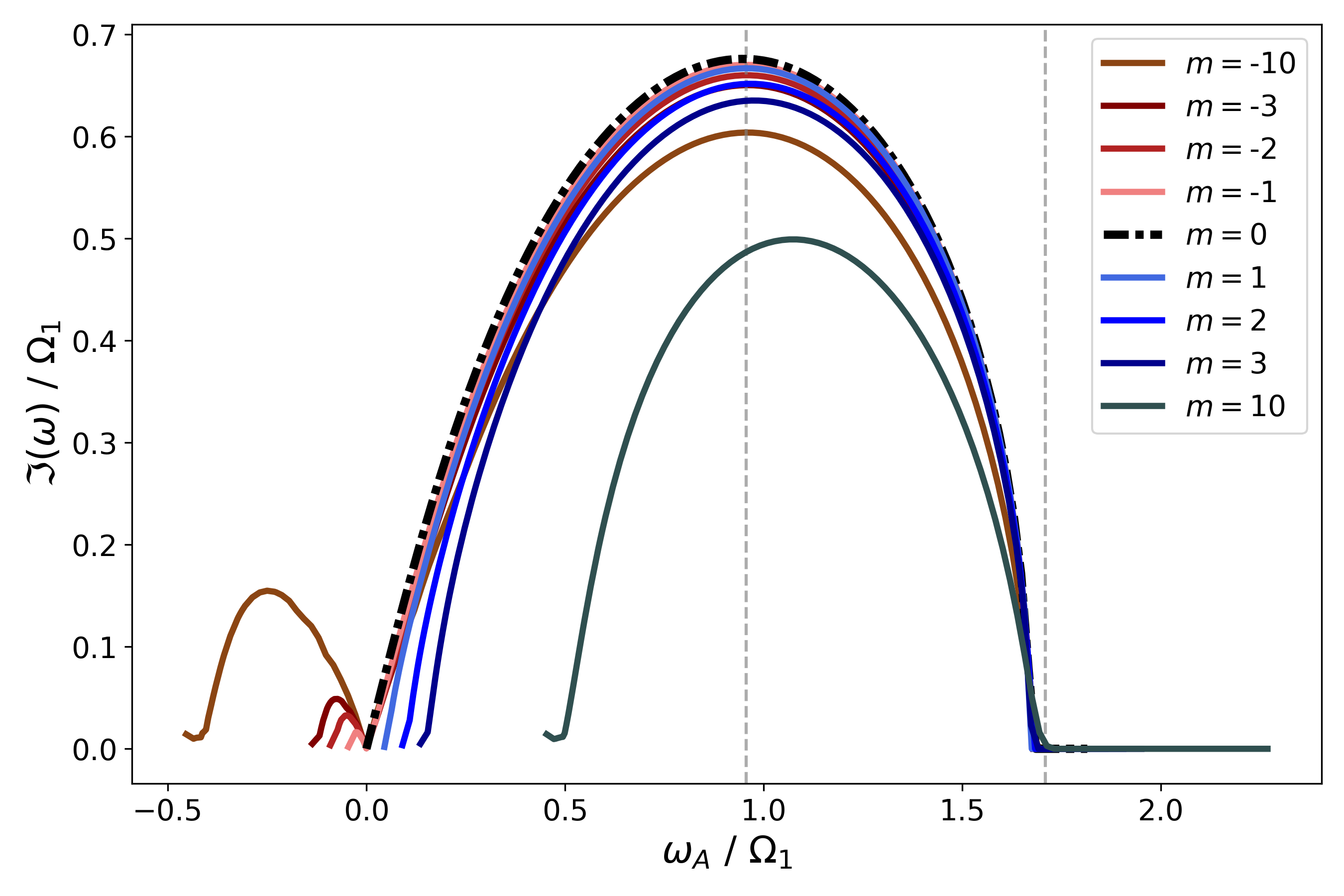}
    \caption{Alternative view of Fig.~\ref{fig:parameter_k}, where the inner SARI growth rates now depend on $\omega_A$ varying through $k$. Instability is perfectly governed by the criterion $0<\omega_A^2 \lesssim 3\Omega_1^2$. Most growth rates peak at around $\omega_A \approx \sqrt{15/16}\ \Omega_1$.}
    \label{fig:parameter_omegaA_k}
\end{figure}

The validity of the instability criterion is made strikingly clear when we plot the growth rates as a function of $\omega_A(r_1)$, as in Fig.~\ref{fig:parameter_omegaA_k}. It contains the same results of Fig.~\ref{fig:parameter_k} where $k$ is varied, but without the outer SARIs. Stability is indeed approximately limited to the region where $\omega_A(r_1) < \sqrt{3} \Omega_1$ and it is clear how the $m<0$ SARIs have access to the whole unstable range, and are stabilised when the Alfvén frequency moves through zero. Note how the $m>0$ SARIs cannot access the whole range of unstable Alfvén frequencies. The maximum growth rates are indeed approximately reached for $\omega_A(r_1) = \sqrt{15/16}\ \Omega_1$, as predicted by Eq.~\eqref{eq:numax_wavenumbers}, but this value shifts depending on $m$. The theoretical maximum growth of $0.75\Omega_1$ is, however, not attained. The slight differences for $\omega_{A\text{max}}$ and the maximum growth rate depending on $m$ are partly explained by the fact that these inner SARI growth rates are evaluated not at the exact Doppler radius where their eigenfunctions peak, but at the inner radius instead. Finally, note that the equilibrium for Fig.~\ref{fig:parameter_omegaA_k} has $v_{A1} \sim \epsilon \ll 1 \sim r_1\Omega_1$ and hence is in the super-Alfvénic regime. This is not in contradiction with Fig.~\ref{fig:parameter_omegaA_k}: indeed, although instability extends to wavenumbers $k$ where $\omega_{A1} > \Omega_1$, the rotation frequency in the local Doppler frame $\omega_{A1} < m\Omega_1$ is still super-Alfvénic. 

The outer SARI growth rates (not shown) obey a similar dependency on $\omega_A(r_2)$ when scaled by $\Omega(r_2)$. Note also that the condition $\omega_A^2(r_2) < 3\Omega^2(r_2)$ for outer SARI instability is more restrictive than that for the inner SARIs, as the right-hand side of the inequality decreases faster than the left-hand side. This explains why outer SARI modes are stabilised more easily compared to inner SARIs. We also noted in the previous Sections that outer SARIs have growth rates three times lower than those of inner SARIs. When the outer SARIs are, however, scaled by the \textit{local} rotation frequency $\approx \Omega(r_2)$ instead of $\Omega_1$, this discrepancy disappears. Indeed, $\Omega(r_2) = \Omega_1 r_2^{-3/2} \approx \Omega_1 / 3$ if $\delta = 1$. Relative to the local frame of reference, the outer modes are hence equally fast-growing over one rotation period, with maximum growth rates up to $0.7 \Omega$ in our numerical results. The fact that the outer SARIs are more easily stabilised than the inner modes gives two possibilities for the evolution of the entire spectrum when an equilibrium parameter is varied: (1) the quasi-continuum first collapses to a point as the overlap between the Alfvén continua decreases, where it might detach from the real axis. The inner and outer branches then merge into one branch of discrete modes, or the outer branch is stabilised before that happens; or (2) the outer branch is stabilised before the quasi-continuum disappears, and the quasi-continuum still exists as long as the overlap of the continua is large enough. In either case, the remaining branch of inner modes is reminiscent of the unstable sequence of MRI modes, but it contains an infinite number of modes clustering to an internal edge of one of the continua $\Omega_A^\pm$ \citep{GK22}. It, too, eventually disappears to the real axis when it is stabilised.

To compare the predictions to the actual values of $k_\text{cut-off}$ in Figs.~\ref{fig:parameter_k}--\ref{fig:parameter_beta}, we evaluate Eq.~\eqref{eq:instability_criterion} at the local Doppler radius $r_*$ where $\sigma = \Omega_0(r_*)$, and the MRI growth rates at the inner boundary $r_1$. These predictions are reasonably accurate when compared to the actual growth rates of Fig.~\ref{fig:parameter_k}: e.g. the $m=-3$ SARI is predicted to have $k_\text{cut-off} = 411$ while in reality $k_\text{cut-off}=400$, and the value of $k$ where the inner mode is locally stabilised is predicted to be $k_\text{stab}=33$ while in reality it happens as $k_\text{stab}=30$. When keeping $k$ fixed in Figs.~\ref{fig:parameter_mu} and \ref{fig:parameter_epsilon}, it is the orientation or strength of the magnetic field that determines whether Eq.~\eqref{eq:instability_criterion} is satisfied. The Alfvén frequency in terms of the equilibrium parameters is given by $\omega_A^2(r) = \left( m\mu_1/r + k \right) \epsilon\left(1+\mu_1^2\right)^{-1/2} r^{-1/2}$. Hence, the instability criterion predicts that, once the field gets stronger, large values of $k$ are only allowed when the toroidal component is significant; this is exactly the difference between Fig.~\ref{fig:parameter_mu} (a) and (b), which both have $k=50$. The instability criterion again explains the ordering of the cut-off vertical field obtained for Case (b): if $m<0$, $B_\theta$ can become smaller compared to $B_z$, and vice-versa for $m>0$. For Case (a), there is no cut-off as $k$ is below the critical value $k_\text{cut-off}$ for a purely poloidal field. For a fixed orientation and wavenumber, instability is only possible in a range of field strengths where the upper limit of Eq.~\eqref{eq:instability_criterion} is satisfied. The ordering of $m<0$ and $m>0$ mode critical field strengths in Figs.~\ref{fig:parameter_epsilon} (a) and (b) is again explained from this criterion, except for the high $v_{A1}$ in Case (b). Interestingly, Eq.~\eqref{eq:instability_criterion} leaves the option for MRI/SARI in a range of wavenumbers to be `switched off' at intermediate pitch angles $\varphi$, since $\omega_A$ as a function of $\varphi$ might first increase above the critical threshold of Eq.~\eqref{eq:instability_criterion} and then again decrease. This implies that some high-$m$ co-rotating SARIs are stabilised for $\varphi \in [5^\circ, 25^\circ]$, for example, which means that they are unstable for an almost purely vertical field or when a significant azimuthal field is present. Finally, a visualisation of $\omega_A$ as $\mu_1$ or $\epsilon$ is varied gives almost the same results as those in Fig.~\ref{fig:parameter_omegaA_k}.

\subsubsection{Growth rate estimates}
We now investigate the upper bound to the SARI growth rates by evaluating Eq.~\eqref{eq:numax_GK22} at the local Doppler corotation radius, where $\sigma = \Omega_0(r_*)$, which is also where the eigenfunctions peak. Expression~\eqref{eq:numax_GK22} for the maximal growth rate varies with $r$ and is usually radially decreasing for our accretion disk equilibria. This implies that inner SARIs (localised around $r_1$) indeed have higher growth rates than outer SARIs (localised around $r_2$), as was found in most of our spectra and parameter scans. There are, however, specific instances where $\nu_\text{max}$ is non-decreasing and the outer SARI can have higher growth rates. Recall for example the cases with vanishing $\omega_A(r_1)$ above, or the $m=-3$ SARI in Fig.~\ref{fig:parameter_beta}. In the second case, $\nu_\text{max}$ increases from its value at $r_1$ up to a global maximum, after which it again decreases to its value in $r_2$, which is larger than that in $r_1$. However, there is no mode corresponding to the maximum of $\nu_\text{max}$ with a growth rate higher than that of the outer mode, so the outer SARI is the most unstable mode. This also points to a second observation: the predicted upper bound for growth rates is never strict. It usually overestimates the growth rates, for both discrete and quasi-continuum modes, but one can nevertheless derive the shape of the quasi-continuum upper edge from it, as in Fig.~\ref{fig:dispersion_prediction}. A possible reason for this discrepancy is that the derivative of the eigenfunctions, $\chi'$, was not taken into account in the derivation of \eqref{eq:numax_GK22}, as noted in \citet{GK22}. We will now use Eqs.~\eqref{eq:numax_GK22} and \eqref{eq:instability_criterion} to explain several observations made earlier.

To fully confirm the finding of Fig.~\ref{fig:parameter_mu_and_k} that a purely toroidal field allows for arbitrarily large unstable wavenumbers $k$, we extend the numerical results with predictions using Eq.~\eqref{eq:numax_GK22}. Indeed, as $\varphi\rightarrow 90^\circ$, the values of $k_\text{max}$ and of $k_\text{cut-off}$ increase dramatically: the growth rates for $\varphi = 89.99^\circ$ are predicted to peak at $k=2\times10^5$, and for $\varphi = 88.999^\circ$ at $k=1.5\times10^6$; the most unstable modes can become arbitrarily localised in the vertical direction. Here, the distinction with the $m=0$ MRI becomes clear: we perform the same \texttt{Legolas} runs as in Fig.~\ref{fig:parameter_mu_and_k} but now for the MRI, extended with analytical predictions for very large $k$. Unsurprisingly, the $m=0$ MRI also shows increased localisation as $\varphi \rightarrow 90^\circ$ because the poloidal component gets weaker and hence larger $k$ are allowed. However, whereas the SARI retains finite growth rates for small $k$ as $\varphi\rightarrow 90^\circ$, the MRI is stabilised in the limit: $k_\text{max}$ moves to arbitrarily large values, but for values of $k \leq 1000$, the growth rates become arbitrarily low. In contrast to the SARI modes, the $m=0$ MRI has no way to couple to the toroidal field, and once the poloidal component becomes negligible, it is stabilised.

Expression~\eqref{eq:numax_GK22} for $\nu_\text{max}$ also explains why $\pm m$ SARIs have coalescing growth rates for $\varphi\rightarrow0^\circ$ or $\varphi\rightarrow 90^\circ$ in Fig.~\ref{fig:parameter_mu} for both Case (a) and Case (b). Indeed, for a purely toroidal field, $\nu_\text{max}$ only depends on $m^2$ and for a purely poloidal field, $m$ completely disappears from the expression (but evaluating $\nu_\text{max}$ at the proper resonance location $r_*$ for each mode does produce the observed dependence on $m$). Physically, a purely vertical or toroidal field induces no preferential (co- or counter-rotating) direction for waves to grow in. As mentioned in Sect.~\ref{sec:magnetisation}, Fig.~\ref{fig:parameter_epsilon} (b) shows a peculiar ordering of the critical field strengths $v_{A1\text{cut-off}}$, which decrease as $m$ increases from $m=-3$ to $10$, but the $m=-10$ SARI has a smaller range of unstable field strengths. From Eq.~\eqref{eq:numax_GK22}, we find the following behaviour of the instability as $m$ varies: for $m>0$, the range of unstable field strengths decreases and disappears in the limit $m\rightarrow\infty$, since then $\omega_A$ becomes high. For $m<0$, the critical field strength first increases to a maximum, then decreases and eventually vanishes too in the limit $m\rightarrow -\infty$. When the field has a dominant toroidal component, this switchback occurs for lower $m$ than when the field is dominantly poloidal.

We conclude that, in general, Eq.~\eqref{eq:numax_GK22} correctly predicts the general trend of growth rates in weakly magnetised, thin disks, the parameter regime for which it was derived. Outside of this regime, it breaks down: when the assumption $k^2r^2 \gg m^2$ is violated in Fig.~\ref{fig:parameter_k} all growth rates nearly vanish as $k\rightarrow0$ but the predicted growth rates remain finite; for $\beta < 1$ in Fig.~\ref{fig:parameter_beta}, growth rates are not at all correctly predicted, and the $m=\pm 10$ SARIs behave in a distinct way. It is clear that in the compressible regime, one has to resort to numerical means as done in this work. Equation \eqref{eq:numax_GK22} does not replicate the decreasing maximum growth rates as $|m|$ increases, which is observed above in most figures.

\subsubsection{Global vs local approach}
The global eigenvalue problem as in \citet{keppens2002} has often been in the shadow of the local WKB approximation because of the involved methods required for its analysis. However, the interaction of nearby singularities of the governing ODE is exactly what is responsible for the special properties of non-axisymmetric modes. We briefly show in this section that local dispersion relations are unable to correctly predict both discrete and quasi-continuum SARI modes. In Appendix \ref{sec:appendix_dispersion_blokland}, we show how such a dispersion relation can be derived from the global formulation using a WKB approach \citep{keppens2002,blokland05}, and how some recent dispersion relations obtained directly from the linearised MHD equations also follow from the global formulation \citep{das18,EP22}. 

\begin{figure}
    \center
    \includegraphics[width=0.6\linewidth]{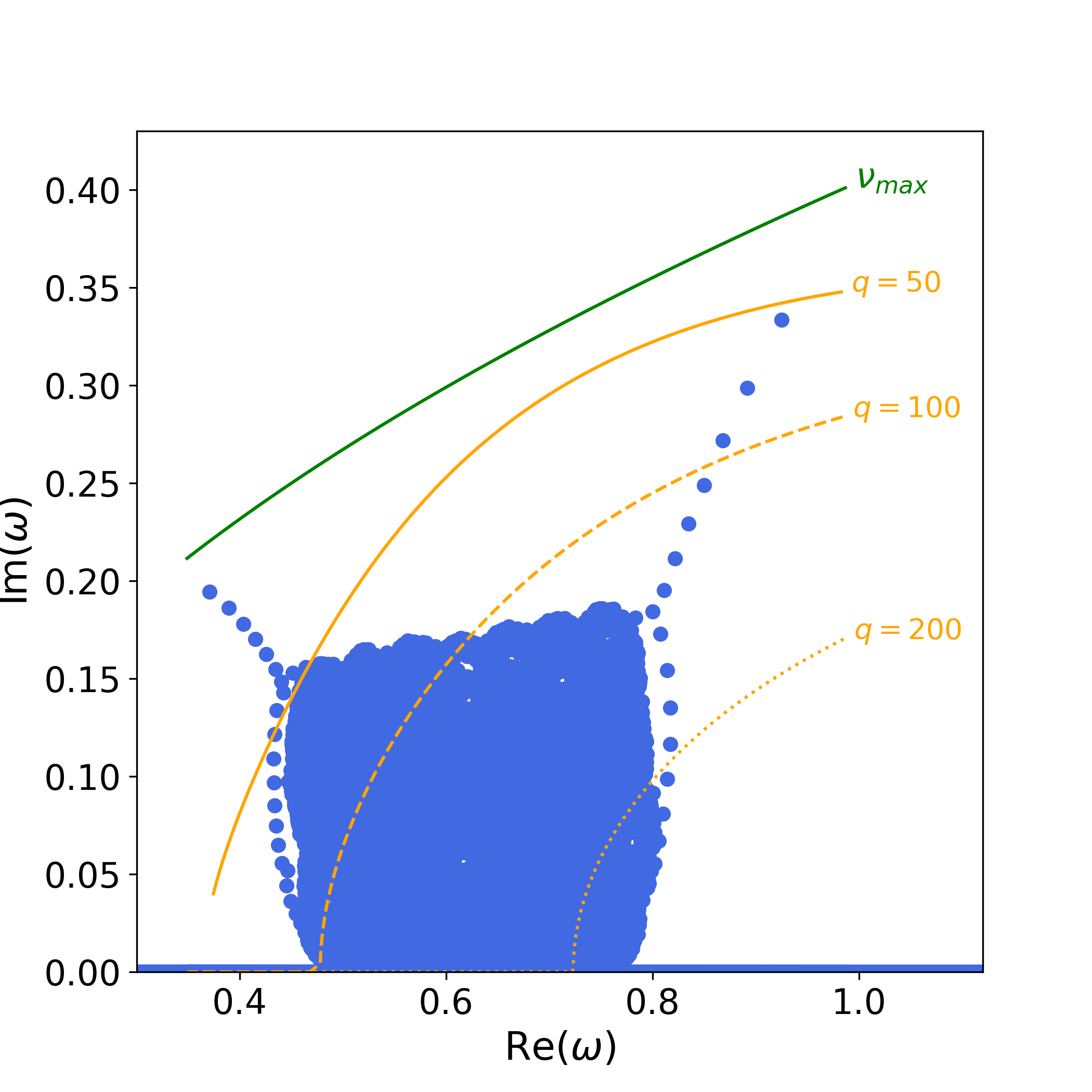}
    \caption{The $m=1$ quasi-continuum of Fig.~\ref{fig:QC_resolution} filled using shift-invert. The prediction of $\nu_\text{max}$ of Eq.~\ref{eq:numax_GK22} (green) is compared to the fastest-growing modes from a WKB dispersion relation \eqref{eq:dispersion_sixth} at three values for the radial wavenumber $q$. Clearly, a local model fails to predict the existence of quasi-modes, and that of the outer SARI.}
    \label{fig:dispersion_prediction}
\end{figure}

Solving a local WKB dispersion relation requires a radius $r$ where all equilibrium parameters should be evaluated, and a radial wavenumber $q$ in addition to $k$ and $m$. \citet{blokland05} showed that a sixth-order local dispersion relation can exactly replicate MRI eigenfrequencies if $r$ and $q$ are derived self-consistently from the peak radius and radial wavelength of an exact eigenfunction of the global formulation. We now evaluate this dispersion relation over the whole radial domain to obtain the growth rates of the most unstable modes, and consider several values of $q$. In Fig.~\ref{fig:dispersion_prediction}, a range of predicted eigenfrequencies from the sixth-order dispersion relation \eqref{eq:dispersion_sixth} is plotted for three fixed values of $q$ (orange). For the most unstable inner SARI, this leads to a good approximation ($q\approx 50$ is the value obtained from the eigenfunction). The same cannot be said for the quasi-continuum SARI modes: since the eigenfunctions for modes with growth rate $\nu$ are very similar in shape but shifted horizontally according to the local Doppler resonance, a constant value of $q$ would have to correspond with a nearly horizontal band of quasi-continuum modes. However, as Fig.~\ref{fig:dispersion_prediction} shows, the maximal growth rates are quickly decreasing towards the outer edge of the disk and the dispersion relation predicts a stabilisation of all modes for all three radial wavenumbers. Note that the larger wavenumbers $q$, which should correspond to radially localised modes, perform worse in predicting ranges of unstable modes, and even start missing out the entire Doppler range. Indeed, also in the local dispersion relation of \citet{terquem_papaloizou}, the MRI is stabilised for $q \gg k$. The WKB analysis assumes radially oscillating solutions over the whole domain, whereas SARI eigenfunctions are radially localised. MRI eigenfunctions can fill the entire radial domain and hence are better suited for such an approximation. In particular, the dispersion relation gives no correct predictions for the outer SARI growth rates. The local WKB approximation is hence unable to replicate the complex interactions of the two singularities, which are crucial in the analysis of SARI modes. In contrast, the green line in Fig.~\ref{fig:dispersion_prediction} representing $\nu_\text{max}$ \eqref{eq:numax_GK22} obtained through global means, provides an upper bound for the discrete SARIs as well as for the quasi-continuum modes, and correctly predicts the radial decrease of the maximum growth rates.

\subsection{Extremely localised modes in global spectra} \label{sec:param_spectra}

In Sect.~\ref{sec:MRI_SARI} and \ref{sec:QC}, we have shown typical MRI and SARI spectra (Fig.~\ref{fig:MRI_SARI}), as well as unstable 2D continua filled with quasi-modes (Fig.~\ref{fig:composite_QC}). We then proceeded in Sect.~\ref{sec:param_study} to calculate the growth rates of the most unstable MRI and SARIs over a range of parameters, showing how the inner SARIs usually dominate over the outer SARIs and how modes can be stabilised by locally vanishing $\omega_A$. However, not just the fastest-growing modes are of interest in this study: the presence of quasi-continua provides a source of unstable modes with localised eigenfunctions all over the disk, as shown in Sect.~\ref{sec:QC}. We set forth to calculate a quasi-continuum of large-$m$ SARIs that are extremely localised (but global in nature). 

In principle, a SARI quasi-continuum can occur whenever $\Omega_A^+$ and $\Omega_A^-$ overlap, but the height that it might reach can be severely limited, requiring a high resolution ($n_\text{grid} \geq 1000$) to adequately resolve its edges. This depends on the ratio $|k|/|m|$, which appears in the exponent of exact quasi-continuum solutions \citep{GK22}. As this ratio increases, the associated complementary energy greatly decreases and as a result, the quasi-continuum `grows'. These two conditions are summarised in Eq.~(91) of \citet{GK22}, which we repeat here:
\begin{equation} \label{eq:QC_condition}
    1 \ll \frac{|k|}{|m|} < \frac{ \epsilon^{-1}\sqrt{1+\mu_1^2}(1-\frac{5}{8}\epsilon^2\beta)(1-(r_1/r_2)^{3/2}) - \text{sgn}(m)\mu_1(1+(r_1/r_2)^{3/2}) }{1+(r_1/r_2)^{1/2}},
\end{equation}
where the upper bound makes sure that the Alfvén continua overlap for the self-similar radial profiles in this set-up.

We use Eq.~\eqref{eq:QC_condition} and the results of Sect.~\ref{sec:param_study} to obtain a SARI quasi-continuum for a high azimuthal mode number $m=-10$. The two conditions need to be balanced: the ratio $|k/m|$ must be high, so a $k \gg |m|$ is needed -- for Fig.~\ref{fig:composite_QC}, $k/|m| = 50$ -- but at the same time the instability has a cut-off with $k$ when the magnetic tension of the vertical field component stabilises the mode, as seen in Sect.~\ref{sec:vertical_localisation}. Additionally, the overlap between $\Omega_A^\pm$ strongly depends on $\epsilon^{-1}$, which needs to be large. A high-$m$ quasi-continuum hence prevails in weak fields, where large $k$ are allowed for instability. Alternatively, a stronger toroidal component moves $k_\text{cut-off}$ to larger values as shown in Sect.~\ref{sec:field_orientation} and thus can also allow large $k$. An additional difficulty is that, for strongly overlapping $\Omega_A^\pm$, SARI resonances are close by and the modes are radially very localised between those resonances, requiring a high resolution for resolving the eigenfunctions. All in all, we use the equilibrium parameters $\mu_1 = 10$, $\beta = 10$, $\epsilon=0.01$, $\delta = 1$ and $k=400$ so that $k/|m| = 40$. 

\begin{figure}
    \plotone{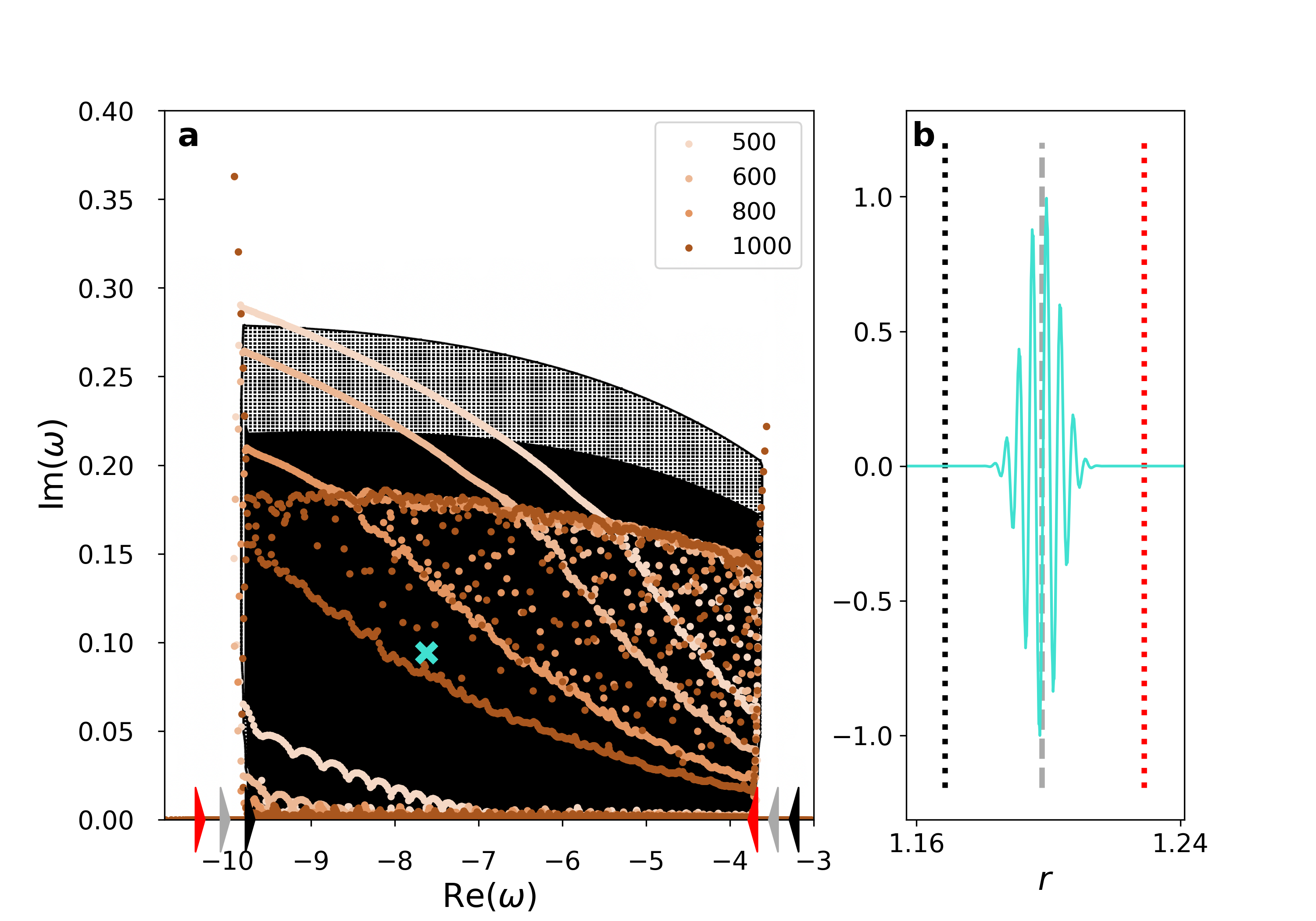}
    \caption{A large-$m$ ($-10$), large-$k$ ($400$) quasi-continuum with radially very localised eigenfunction $\Re(v_r)$ (cyan) ($\omega=-7.62+0.09i$) as a result of the large overlap between $\Omega_A^\pm$. Increasing grid resolution finds edges that match $W_\text{com}=10^{-12}$ (black). Equilibrium parameters: $\epsilon=0.01$, $\mu_1=10$, $\beta=10$, $\delta=1$.}
    \label{fig:QC_m-10}
\end{figure}

Figure~\ref{fig:QC_m-10} shows the results of four QR-runs with increasing resolution that gradually find the true edges of a quasi-continuum for $m=-10$, overlaid on a quasi-continuum calculated with \texttt{ROC} at two values of $W_\text{com}$. Both the left and upper edges of the quasi-continuum are only fully resolved at a \texttt{Legolas} resolution of $n_\text{grid}=1000$. The latter has a value of $\nu \approx 0.2$, which is comparable to the results of Fig.~\ref{fig:composite_QC}. The scattered modes found within the quasi-continuum bounds are part of dense clouds of eigenvalues when applying Arnoldi-shift-invert to locations inside the 2D region (not shown). Furthermore, the eigenfunctions, shown in Fig.~\ref{fig:QC_m-10} (b), again resemble wave packages localised between the Alfvén resonances, which are now minimally separated (by only $\sim 1\%$ of the radial domain!). Hence, the criteria from Sect.~\ref{sec:QC} to classify this \texttt{Legolas} spectrum as a quasi-continuum have been met. Using a modified version of \texttt{ROC} that enables shooting between the very close-by Alfvén resonances rather than over the whole domain, we see that the eigenfunctions oscillate heavily for $\nu\rightarrow 0$ and that they fill the whole region between the resonances. All together, we have modes that are highly localised in all directions: azimuthally with $m=-10$, longitudinally with $k=400$ and strongly oscillating radially in a small portion of the accretion disk, far away from the boundaries. Hence, these modes are truly localised, even though they are global in nature. It is hence not unthinkable that they have shearing box analogs, where only a tiny radial extent still produces overlapping continua that generate SARI-like modes \citep{matsumoto_tajima95,noguchi2000}.

\section{Discussion and conclusions}

The first part of this work considered the SARI quasi-continua discovered by \citet{GK22}. We independently confirmed their existence, noting that they are truly indistinguishable from true normal modes by any numerical code - and physical plasma, for that matter. Although the mathematical treatment of these modes is quite novel, hints of their existence were already present in earlier works: the non-axisymmetric spectra of \citet{keppens2002} show large collections of modes that are reminiscent of the spectra in Fig.~\ref{fig:QC_resolution}, where the QR-algorithm finds many scattered modes inside a quasi-continuum region. An earlier example are the modes excited in two-dimensional simulations by \citet{terquem_papaloizou}, which have some resemblance with the quasi-mode eigenfunctions and move at the local Doppler speed. More analysis is needed to determine whether they are truly like the wave packages seen in Sect.~\ref{sec:QC}. In any case, it is clear that a global treatment of the problem is essential to capture these effects: the nearness of singularities in the overlapping continua lies at the origin of the problem, and this cannot be obtained through a local analysis. With the definition of quasi-modes having a tiny but finite complementary energy on a firm physical and mathematical basis, we presented how quasi-modes can be identified using modern MHD spectroscopy with a tool like \texttt{Legolas}. The \texttt{ROC} and \texttt{Legolas} codes can be operated in tandem, so that the predictory value of the latter can be tested against the explicit $W_\text{com}$ calculations of the former, as was demonstrated in Fig.~\ref{fig:QC_m-10}. It will be interesting to establish whether a shearing box equivalent of SARIs exists in our formalism, resembling the radially localised non-axisymmetric modes found by \citet{matsumoto_tajima95} and \citet{noguchi2000}. Global high-$m$ modes with eigenfunctions that are adequately localised could perhaps even survive in the small radial variation of a local box.

In the second part, we performed a comprehensive study of instability growth rates in a wide range of equilibria, encompassing accretion disks with weak to strong, vertical to azimuthal, and subthermal to suprathermal magnetic fields. The stabilisation of modes at various values of $m$ was investigated for the most unstable inner and outer SARIs and the MRI. It was shown that stability is indeed governed by a condition \eqref{eq:instability_criterion} on the Alfvén frequency that follows from analytical theory, which was used to explain growth rates for weakly magnetised, thin disks. A recent upper bound for the growth rates \eqref{eq:numax_GK22} explained why the outer SARI growth rates behave like a scaled-down version of their inner equivalents. We also concluded that the $m<0$ counter-rotating SARIs play an important role in opening up more regimes for instability, significantly extending the range of unstable equilibria at high $k$, as shown in Fig.~\ref{fig:parameter_k} and Fig.~\ref{fig:parameter_epsilon} (b). Our findings for non-axisymmetric modes and their implications are summarised below. In particular, the maximal values of the numerically calculated growth rates are $\sim 0.6 \Omega_1$, in correspondence with the global analysis of \citet{terquem_papaloizou} and \citet{begelman22}, while \citet{BH92b} find growth rates of only several percent of $\Omega_1$ in a local shearing sheet approximation. This again highlights the importance of using a global method when dealing with non-axisymmetric modes. 

\textit{Localisation --} in a thin, weakly magnetised disk, the fastest-growing mode is found at intermediate to small wavelengths, with both $k_\text{max}$ and $k_\text{cut-off}$ increasing with increasingly negative $m$ as instability depends on $\omega_A$, a conclusion also drawn by \citet{begelman22}. For equilibria with a sizeable toroidal field, the highest growth rates are found at large $k$ \citep{BH92b,ogilvie_pringle}. This implies that the essence of the instability is still captured in this approximation of an infinitely long cylinder, which requires the assumption of small vertical wavelengths compared to the scale height, and the modes found represent modes that are localised around the midplane of a 3D accretion disk \citep{BH98}. In contrast, in models of accretion tori and 3D shearing boxes, unstable modes near the vertical edges of the disk are found \citep{BH98}, as well as strongest growth for the MRI at wavelengths about the size of the vertical scale height $H$ \citep{mamatsashvili13}. 

Additionally, we showed in Sect.~\ref{sec:param_spectra} that large-$m$ modes have significant maximum growth rates, usually lower than those of the MRI but sometimes comparable. However, given the vast range of accessible wavelengths for non-axisymmetric instabilities, even those with growth close-to-optimal can dynamically outperform the axisymmetric MRI \citep{mamatsashvili13}. In any case, the obtained growth rates are an order of magnitude higher than those applicable to the shearing sheet model by \citet{BH92b}, as was also remarked by \citet{curry_pudritz96}. The possibility of a locally vanishing $\omega_A$ locally stabilises SARI modes. We also explicitly showed that quasi-continua do exist in configurations with large $m$ and $k$ (Fig.~\ref{fig:QC_m-10}). Quasi-continuum SARI modes with such large wavenumbers are hence also very relevant to shearing box simulations, where the simulation domain is radially, vertically, and azimuthally localised \citep{lesaffre2009,bai_stone13,hirai2018}. The quasi-modes resulting from these spectra can corotate anywhere in the radial domain, making them ideal candidates for instability in both local Cartesian and global simulations.

In practice, however, large-$k$ and large-$m$ modes can only be realised as long as the scales involved are above the resistive and viscous length scales. The effect of these non-ideal terms will generally be to quench the instability, particularly on smaller scales \citep{BH91,BH98,kitchatinov2010}. We plan to further investigate the effect of these diffusive terms on the MRI/SARI, especially on localised quasi-continuum SARI modes. Since the inclusion of non-ideal MHD effects breaks the up-down symmetry of MHD spectra \citep{GKP19}, it will be interesting to see how the quasi-continuum regions are modified in the structure of the spectra. We note that \texttt{Legolas} can analyse a number of non-ideal effects in combination, and thus can explore visco-resistive, or Hall and ambipolar effects on the ideal MHD spectrum.

\textit{Magnetic field --} We can conclude from Sect.~\ref{sec:field_orientation} and \ref{sec:magnetisation} that a strong poloidal field inhibits large-$k$ instabilities, while a strong toroidal field inhibits large-$m>0$ instabilities: the magnetic tension produced by these small-wavelength modes would be too high \citep{BH98}. As a result, if the field is predominantly poloidal and strong, the MRI will be stabilised at small $k$ while negative $m$ modes can still have significant growth rates for large $k$ if $\omega_A$ satisfies criterion \eqref{eq:instability_criterion} \citep{begelman22,begelman23}, as was also confirmed by Fig.~\ref{fig:parameter_mu}. In particular, counter-rotating SARIs have one value of $m$ where the unstable range of field strengths is maximally extended, as in Fig.~\ref{fig:parameter_epsilon} (b). This implies that strong, poloidal fields could be unstable to extremely high-$m$ modes beyond the capabilities of our linear (and non-linear) codes: their eigenfunctions would be constrained to a tiny fraction of the domain. We also confirmed that for a purely toroidal field, a \textit{`complete breakdown of the most rapidly growing local modes'} \citep{BH98} does indeed occur: the $k_\text{max} \rightarrow \infty$ for a purely toroidal field predicted by \citet{ogilvie_pringle} quickly moves to much smaller values for even a seed poloidal field, as was noted in Fig.~\ref{fig:parameter_mu_and_k}. Finally, in our exploration of the plasma-beta, we found that a dominant toroidal field has low growth rates for $\beta < 1$. This was also found by \citet{das18} at intermediate field strengths, but they identified a new hybrid Alfvén/slow instability at stronger fields. It is unclear if a similar instability appears for our $m=10$ SARI in Case (a). According to \citet{van_der_swaluw05}, these instabilities must be more MRI-like than convection-like, since the parameter range explored here corresponds to thin disks ($\sqrt{p_1} \leq 0.6$).

Not only the fastest-growing modes are of importance for stability: the linear evolution is governed by an entire MHD spectrum of eigenmodes. We found that the outer SARI branch is usually stabilised first, implying that the quasi-continuum between the discrete SARI branches can detach from the real axis and disappear before that, or continue to exist until the inner SARIs are also stabilised. It should be noted that we did not cover SARI quasi-continua for a dominant poloidal field, but these do exist. We confirmed this for $m=1$ at weak field strength. The Appendix covers the basic theory needed for the global formulation of the eigenvalue problem, from which a local dispersion relation can be derived. We show the connection between the general dispersion relation from \citet{blokland05} and recent work by \citet{das18} and \citet{EP22}. By comparing solutions to this dispersion relation with a SARI spectrum in Fig.~\ref{fig:dispersion_prediction}, we highlighted that the local approximation is insufficient to describe outer and quasi-continuum SARI eigenmodes. 

\begin{acknowledgments}
    N.B. is supported by an FWO Flanders fellowship titled \textit{``Magnetoseismology of accretion disks: towards the full spectrum for turbulent disks"} - grant number 11J2622N. R.K. acknowledges support by the ERC Advanced Grant PROMINENT from the European Research Council (ERC) under the European Union’s Horizon 2020 research and innovation programme (grant agreement No. 833251 PROMINENT ERC-ADG 2018), the C1 project TRACESpace funded by KU Leuven, and FWO project G0B4521N.
\end{acknowledgments}

\appendix

\section{Frieman-Rotenberg formalism} \label{sec:appendix_spectral_web}

The MHD stability of a moving fluid is most easily derived using a Lagrangian description, a method pioneered by \citet{frieman-rotenberg1960}. This formalism is completely general, where the momentum equation is written in terms of the Lagrangian perturbation of a fluid parcel, $\bxi$. In the case where the equilibrium is only varying radially, the formalism can be reduced to a single second-order ODE for the Fourier coefficient $\hat{\bxi}$ and complex frequency $\omega$ \citep{hameiri1981}. This ODE is solved using a shooting method, or the related, powerful Spectral Web method \citep{goedbloed2018spectralI,goedbloed2018spectralII,GK22}, or can be treated analytically under certain assumptions. 

In the Frieman-Rotenberg formalism \citep{frieman-rotenberg1960}, the Lagrangian displacement $\bxi$ is related to the perturbed velocity through the equation
\begin{equation} \label{eq:xi_and_v}
    \boldv = \left(\frac{\partial}{\partial t} + \bvn\cdot\nabla\right)\bxi - \bxi\cdot\nabla\bvn,
\end{equation}
where $\bvn$ is the background flow field. Note that, in the absence of flow, the usual relationship $\partial \bxi/\partial t = \boldv$ is obtained. By defining a generalised force operator,
\begin{align} \label{eq:force_operators}
    \mathbf{G}(\bxi) & = \mathbf{F}(\bxi) + \nabla\left(r\Lambda\xi_r\right)   + \nabla\cdot\left(\bxi\rho\bvn\cdot\nabla\bvn\right) - \rho \left(\bvn\cdot\nabla\right)^2\bxi,
\end{align}
which contains the static MHD force operator $\mathbf{F}$ as defined in \citet{GKP19} with additional contributions from the background flow, the equation of motion can be written in terms of $\bxi$:
\begin{equation} \label{eq:eq_of_motion}
    \frac{\partial^2 \bxi}{\partial t^2} + 2\rho \bvn\cdot\nabla\frac{\partial \bxi}{\partial t} - \mathbf{G}(\bxi) = 0 ,
\end{equation}
From this equation, the spectral equation can be obtained by applying a Fourier definition to $\bxi$:
\begin{equation} \label{eq:spectral_eq}
    \left[\mathbf{G} - 2\omega U + \rho\omega^2\right]\bxi = 0 ,
\end{equation}
The operator $U = -i\rho\bvn\cdot\nabla$ contains the gradient along the background flow and produces a Doppler shift in the spectral equation. Note that Eq.~\eqref{eq:spectral_eq} is quadratic in the eigenvalue $\omega$. However, contrary to popular belief \citep{EP22}, the operators $\mathbf{G}$ and $U$ themselves \textit{are} self-adjoint (see Chapter 12 of \citet{GKP19} for a proof), making the difficulty of the spectral problem not one of non-self-adjointness but one of a non-linear eigenvalue equation with non-orthogonal eigenfunctions \citep{GK22}. In the Spectral Web method, the notion of a complementary energy is used to define exact modes and quasi-modes. This quantity is calculated from the total energy of an open system \citep{GKP19}:
\begin{equation} \label{eq:complementary_energy}
    W := -\frac{1}{2} \int \bxi^* \cdot \mathbf{G}(\bxi) dV = W^p + W_\text{com} ,
\end{equation}
where $W$ is the total energy, $W^p$ is the potential energy and $W_\text{com}$ is the complementary energy . The former is real and calculated from the symmetric part of the integrand, whereas $W_\text{com}$ can be complex and represents a surface integral over the boundaries. For normal modes, the complementary energy is exactly zero, and so has vanishing real and imaginary parts. These two conditions can be used to construct two paths in the spectral plane on which eigenvalue solutions must lie. Physically, the complementary energy represents the additional energy required to bring a mode into resonance with the time dependence $\exp(-i\omega t)$. True normal modes are natural resonances and hence have vanishing $W_\text{com}$. In the Spectral Web method, non-zero $W_\text{com}$ implies a discontinuity in the total pressure perturbation where the left and right solutions obtained from shooting from the boundaries should match. For quasi-continuum SARIs, this jump is so small that the modes can almost be regarded as normal modes.

The vector equation \eqref{eq:spectral_eq} in $\bxi$ can be reduced to a scalar ODE by replacing gradients in the azimuthal and vertical directions by their Fourier-transformed representations, reducing the equation to one for the Fourier coefficients $\hat{\bxi}$. The second and third components of $\hat{\bxi}$ can then be written in terms of the first component $\xi_r$. These expressions can then be used to write the equation of motion in terms of $\chi = r\xi_r$ alone as a second-order ODE:
\begin{equation} \label{eq:ODE}
    \frac{d}{dr} \left( \frac{N}{D} \frac{d}{dr}\chi \right) + \left( A + \frac{B}{D} + \frac{d}{dr}\left(\frac{C}{D}\right) \right) \chi = 0 ,
\end{equation}
This generalised Hain-Lüst equation was first obtained in the context of accretion disks by \citet{keppens2002} in a different form, and was rewritten by \citet{blokland05} in the form above. It contains function coefficients that are in general quite complicated expressions involving equilibrium quantities, the wavenumbers $m$ and $k$ and the frequency $\omega$. They can be found in \citet{GKP19} and \citet{GK22}, but we will repeat the most important definitions here.

As noted before, the operator $U$ produces a Doppler shift in the frequency. In cylindrical coordinates, the Doppler range is given by
\begin{equation}
    \Omega_0(r) = m\Omega(r) + k v_{z}(r) ,
\end{equation}
so that the Doppler-shifted frequency becomes
\begin{equation} \label{eq:doppler_shift}
    \omegat = \omega - \Omega_0 .
\end{equation}
Note that the Doppler range has a radial dependence and vanishes for axisymmetric perturbations in our equilibrium without vertical flow.

The singular solutions of Eq.~\eqref{eq:ODE} are produced by the coefficients $N$ and $D$. Actually, the latter function only leads to \textit{apparent} singularities, which becomes clear after rewriting Eq.~\eqref{eq:ODE} as a system of first-order ODEs \citep{blokland05}. The \textit{genuine} singularities, resulting from the zeroes of $N$, can be obtained explicitly:
\begin{align} \label{eq:continua}
    N(r) = \frac{\tilde{A}\tilde{S}}{r}; \qquad \tilde{A} & = \rho (\omegat^2 - \omega_A^2) = \rho (\omega - \Omega_A^+)(\omega - \Omega_A^-);                               \\
    \tilde{S}                                             & = \rho(\gamma p + B^2)(\omegat^2 - \omega_S^2) = \rho(\gamma p + B^2)(\omega - \Omega_S^+)(\omega - \Omega_S^-).
\end{align}
The four singular ranges are dubbed forward and backward Alfvén and slow continua, as the frequencies lie in a continuum band of Alfvén frequencies $\omega_A^2$ and slow frequencies $\omega_S^2 = \omega_A^2 \ \gamma p/(\gamma p + B^2)$ shifted by the Doppler range $\Omega_0$. The eigenfunctions corresponding to these continuum frequencies are non-square integrable functions with a singularity at the location of resonance with the continuum, i.e. where $\Re(\omegat) = \omega_{A,S}(r_\text{res})$.

In the incompressible limit, where $\gamma\rightarrow\infty$ and hence $\omega_S^2 \rightarrow \omega_A^2$, the spectral ODE takes a slightly simpler form \citep{GK22},
\begin{equation}
    \label{eq:ODE_incompressible}
    r \frac{d}{dr}\left( \frac{\tilde{A}}{r h^2} \frac{d\chi}{dr} \right) - \left(\tilde{A} + \Delta - \frac{4k^2 P^2}{h^2 \tilde{A}} - 2mr\frac{d}{dr}\left(\frac{P}{r^2 h^2}\right) \right)\chi = 0,
\end{equation}
where $h^2 = m^2/r^2 + k^2$. For brevity of notation, we use the same definitions of the equilibrium functions as \citet{GKP19},
\begin{equation} \label{eq:equilbrium_functions}
    \Delta = r \left(\frac{B_\theta^2}{r^2} - \rho\Omega^2\right) + \rho' \frac{GM_*}{r^2}, \qquad
    P = \frac{B_\theta}{r}F + \rho\Omega\omegat,
\end{equation}
where $F = \frac{m}{r} B_\theta + kB_z$ is the Fourier-transformed gradient along the magnetic field.

\section{Dispersion relation} \label{sec:appendix_dispersion_blokland}
In the global formulation above, the eigenfunctions have a radial dependence that obeys the second-order ODE \eqref{eq:ODE}. In the numerical survey in the main body of this work, we found that the eigenfunctions of MRI, discrete SARI, and quasi-continuum modes clearly differ, each having its own typical radial variation and (in)sensitivity to boundary conditions. In the local limit, however, a dispersion relation can be derived assuming a purely sinusoidal radial dependence. This Section connects some recent results on accretion disk dispersion relations and non-axisymmetric instabilities.

In the usual WKB analysis \citep{blokland05}, the Fourier coefficients of the eigenfunctions now have a radial dependence, 
\begin{equation} \label{eq:WKB_chi}
    \chi(r) = \tilde{\chi}(r) \exp\left(i \int_{r_1}^{r} q(s)ds\right),
\end{equation}
where $\tilde{\chi}(r)$ is the amplitude and $q(r)$ represents a radial `wavenumber'. This wavenumber is then assumed to be large compared to the radial length scales of background equilibrium gradients, i.e. $q\Delta r \gg 1$. When the \textit{Ansatz} \eqref{eq:WKB_chi} is substituted in a modified version of Eq.~\eqref{eq:ODE},
\begin{equation} \label{eq:ODE_short}
    \left(f\chi'\right)'+g\chi = 0,
\end{equation}
the functions $\tilde{\chi}$ and $q$ can be determined from the real and imaginary parts of the equation and by using the WKB assumption \citep{blokland05}:
\begin{equation}
    q^2(r) \approx \frac{g(r)}{f(r)}, \qquad \tilde{\chi}(r) \approx \frac{1}{(f(r)g(r))^{1/4}}.
\end{equation}
Finally, a sixth-degree dispersion relation in $\omegat$ is obtained from the first equality \citep{keppens2002}, which is of the form
\begin{equation} \label{eq:dispersion_sixth}
    \omegat^6 + c_5 \omegat^5 + c_4 \omegat^4 + c_3 \omegat^3 + c_2 \omegat^2 + c_1 \omegat + c_0 = 0 ,
\end{equation}
where the coefficients can be found in \citet{blokland05}. It describes the relation between $\omegat$, the wavenumbers $(q,m,k)$, and the equilibrium quantities at every location $r \in [r_1,r_2]$. This dispersion relation contains various local dispersion relations obtained for specific approximate cases, including the original standard weak-field MRI dispersion relation by \citet{BH91} as shown in \citet{blokland05}, and fourth- and sixth-order relations by \citet{kim_ostriker} and by \citet{BH98}, respectively, as demonstrated in \citet{keppens2002}. The sixth-order dispersion relation \eqref{eq:dispersion_sixth} describes any general cylindrical disk equilibrium for any $(m,k)$-combination. Its six solutions consist of four oscillating modes (the Alfvén and fast modes) and a conjugate pair of unstable modes (MRI- or SARI-like modes). It was originally applied to the overstability of the $m=0$ mode, which is well described by this WKB approach.

\subsection{Recent dispersion relations}
With the above approach, a local dispersion relation is self-consistently derived from the global framework of the generalised Hain-lüst equation. However, most dispersion relations in the literature are derived by directly manipulating the linearised MHD equations and applying WKB arguments throughout. The two approaches are mutually consistent, since these relations are all contained in the general description by \citet{blokland05}. In this Section, we explicitly show this for two recent examples. 

\citet{das18} derived a general fourth-order dispersion relation from the linearised MHD equations, incorporating a suprathermal poloidal field, curvature, compressibility, and non-axisymmetric modes (their Eq.~(40)). It is indeed contained in the description by \citet{blokland05} and can be obtained by applying the following approximations to Eq.\eqref{eq:dispersion_sixth}:
\begin{itemize}
    \item assuming the fast frequencies are far away, i.e. $|\omegat| \ll q c_S, h c_S$,
    \item an ordering on the wavenumbers, $q^2 \sim k^2 \gg m^2/r^2 \sim 1/r^2$,
    \item a dominant suprathermal toroidal field component, $ B_\theta  \gg B_z$, $\beta \ll 1$,
    \item neglecting the equilibrium variations of the thermodynamic variables, i.e. setting $ \rho' = p' = 0$.
\end{itemize}
These final two assumptions allow us to simplify the deviation from Keplerian force balance $\Lambda$ as
\begin{equation}
    \Lambda(r) := \frac{\left( p + B^2/2 \right)'}{r} + \frac{B_\theta^2}{r^2} \approx \frac{\left(B_\theta^2\right)'}{2r} + \frac{B_\theta^2}{r^2},
\end{equation}
and after rearranging terms, their dispersion relation (40) is obtained. Note that the first condition implies the WKB assumption $q^2 r^2~\gg~1$. It ensures that curvature terms from radial derivatives of the perturbations are ignored with respect to the background. \citet{das18} compared the solutions of this dispersion relation to solutions of the global eigenvalue problem in the limit of a dominant suprathermal toroidal field. Their main focus was to extend and confirm local predictions by \citet{PP05}, who found that the MRI is suppressed entirely over all wavenumbers for strong enough $v_{A\theta1}$. The global study does not find a complete stabilisation, but rather very low growth rates for $k < k_\text{cut-off}$. However, two instabilities appear for higher toroidal fields that are unstable slow and hybrid slow-Alfvén modes, and have overlapping ranges of unstable wavelengths $k$. These modes are also predicted by the dispersion relation, but their cut-off wavenumbers do not overlap. Our Case (b) for the growth rate variation with $\beta$ (Sect.~\ref{sec:magnetisation}) is somewhat similar to their set-up, but we do not observe a stabilisation of the MRI at a certain field strength, the reason being that while $\beta$ is varied, $v_{A1}$ remains constant and hence it is the sound speed that varies. Note that in all cases these authors assume $m=0$, so there is of course no occurrence of a quasi-continuum.

Recently, \citet{EP22} considered non-axisymmetric instabilities in a cylindrical equilibrium flow relevant for MRI laboratory experiments. The authors similarly arrived at a local dispersion relation including curvature and non-axisymmetric effects. There are, however, several notable differences from the analysis presented in \citet{blokland05} and \citet{das18}.

Perhaps the most distinct difference is that the adopted stationary equilibrium significantly deviates: in \citet{EP22}, balance against the centrifugal force is provided solely by the plasma pressure, since the flow is Keplerian with a currentless (potential) magnetic field, no gravity, and a constant radial density. The pressure -- and hence the disk temperature -- is hence implicitly forced to be radially increasing through Eq.~\eqref{eq:radial_force_balance}, in contrast to the negative power laws used in \citet{blokland05}, \citet{das18} and \citet{GK22}. This setup, where a rotating plasma column is pressure-balanced, may well be more relevant for laboratory experiments than for physical accretion disks. However, their fourth-order dispersion relation Eq. (7) contains various terms that cannot directly be reconciled with any of the aforementioned works. Their dispersion relation is derived from linearised momentum and induction equations, but incompressibility is implicitly assumed and eigenfunctions $\rho_1$, $p_1$, $v_{z1}$ and $B_{z1}$ have been eliminated. The authors argue for introducing a curvature term that is not included in previous works, replacing the radial wavenumber $q$ by $\bar{q} := q + 1/r$. However, the inclusion of such a `curvature' term $1/r$, which relates to the radial derivative in cylindrical coordinates, introduces imaginary terms $\sim iq$ that are not present in other published dispersion relations. In a true WKB approximation, the wavenumber $q$ is not modified because of the assumption $qr \gg 1$, and curvature from the cylindrical geometry is only included in the dispersion relation through the azimuthal components $v_\theta/r$ and $B_\theta/r$. With this approximation, the factors $iq$ are still not eliminated from the dispersion relation. In the derivation of \citet{das18}, the following two assumptions prevent these factors from appearing:
\begin{align}
    P_1    & \approx -B_\theta B_{\theta1}, \\
    B_{z1} & \approx -\frac{q}{k} B_{r1}.
\end{align}
The first assumption results from neglecting $v_{z1}$ and $B_{z1}$ in the vertical momentum equation (justified by Eq. (2.22) of \citet{BH91}), while the second assumption follows from the ordering $q^2 r^2 \sim k^2r^2 \gg m^2$ applied to $\nabla\cdot\mathbf{B_1} = 0$. With the mutual consistency of the \citet{blokland05} and \citet{das18} results, we conclude that the inclusion of terms $\sim iq$ is not justified. \citet{das18} came to a similar conclusion when comparing their dispersion relation to that of \citet{PP05}.

Finally, \citet{EP22} consider evanescent solutions in the radial direction: $\chi(r) \sim \exp(-qr)$. This is inconsistent with the approach in Eq.~\eqref{eq:WKB_chi}, since in this \textit{Ansatz} for $\chi$, the real exponential would just be absorbed in the amplitude $\tilde{\chi}(r)$. An a priori evanescent solution is hence impossible in the \citet{blokland05} formalism. The authors then proceed to derive an ODE for the scaled radial displacement $\chi := r\xi_r$ from the momentum equation, equivalent to Eq.~\eqref{eq:ODE}. However, we are unable to reduce their Eq.~(16) to the equivalent equations from \citet{keppens2002}, \citet{blokland05} and \citet{GK22}. A dimensional analysis already reveals some inconsistencies in the ODE, in addition to a sign error in their Eq.~(14).

\bibliographystyle{aasjournal}
\bibliography{refs}{}

\begin{thebibliography}{}
\expandafter\ifx\csname natexlab\endcsname\relax\def\natexlab#1{#1}\fi
\providecommand{\url}[1]{\href{#1}{#1}}
\providecommand{\dodoi}[1]{doi:~\href{http://doi.org/#1}{\nolinkurl{#1}}}
\providecommand{\doeprint}[1]{\href{http://ascl.net/#1}{\nolinkurl{http://ascl.net/#1}}}
\providecommand{\doarXiv}[1]{\href{https://arxiv.org/abs/#1}{\nolinkurl{https://arxiv.org/abs/#1}}}

\bibitem[{Bacchini {et~al.}(2022)Bacchini, Arzamasskiy, Zhdankin, Werner,
  Begelman, \& Uzdensky}]{bacchini2022}
Bacchini, F., Arzamasskiy, L., Zhdankin, V., {et~al.} 2022, \apj, 938, 86

\bibitem[{{Bacchini} {et~al.}(2024){Bacchini}, {Zhdankin}, {Gorbunov},
  {Werner}, {Arzamasskiy}, {Begelman}, \& {Uzdensky}}]{bacchini2024}
{Bacchini}, F., {Zhdankin}, V., {Gorbunov}, E.~A., {et~al.} 2024, arXiv
  e-prints, arXiv:2401.01399

\bibitem[{Bai \& Stone(2013)}]{bai_stone13}
Bai, X.-N., \& Stone, J.~M. 2013, \apj, 769, 76

\bibitem[{Balbus \& Hawley(1991)}]{BH91}
Balbus, S.~A., \& Hawley, J.~F. 1991, \apj, 376, 214

\bibitem[{Balbus \& Hawley(1992)}]{BH92b}
---. 1992, \apj, 400, 610

\bibitem[{Balbus \& Hawley(1998)}]{BH98}
---. 1998, Reviews of modern physics, 70, 1

\bibitem[{{Begelman} \& {Armitage}(2023)}]{begelman23}
{Begelman}, M.~C., \& {Armitage}, P.~J. 2023, \mnras, 521, 5952

\bibitem[{Begelman {et~al.}(2022)Begelman, Scepi, \& Dexter}]{begelman22}
Begelman, M.~C., Scepi, N., \& Dexter, J. 2022, \mnras, 511, 2040

\bibitem[{Blaes \& Hawley(1988)}]{blaeshawley}
Blaes, O.~M., \& Hawley, J.~F. 1988, \apj, 326, 277

\bibitem[{Blokland {et~al.}(2005)Blokland, Van~der Swaluw, Keppens, \&
  Goedbloed}]{blokland05}
Blokland, J., Van~der Swaluw, E., Keppens, R., \& Goedbloed, J. 2005, \aap,
  444, 337

\bibitem[{Chandrasekhar(1960)}]{chandrasekhar1960}
Chandrasekhar, S. 1960, Proceedings of the National Academy of Sciences, 46,
  253

\bibitem[{Claes {et~al.}(2020)Claes, De~Jonghe, \& Keppens}]{claes2020legolas}
Claes, N., De~Jonghe, J., \& Keppens, R. 2020, \apjs, 251, 25

\bibitem[{Claes \& Keppens(2023)}]{legolas2}
Claes, N., \& Keppens, R. 2023, Computer Physics Communications, 291, 108856

\bibitem[{Curry \& Pudritz(1996)}]{curry_pudritz96}
Curry, C., \& Pudritz, R.~E. 1996, \mnras, 281, 119

\bibitem[{Das {et~al.}(2018)Das, Begelman, \& Lesur}]{das18}
Das, U., Begelman, M.~C., \& Lesur, G. 2018, \mnras, 473, 2791

\bibitem[{De~Jonghe {et~al.}(2022)De~Jonghe, Claes, \& Keppens}]{legolasjordi}
De~Jonghe, J., Claes, N., \& Keppens, R. 2022, Journal of Plasma Physics, 88,
  905880321

\bibitem[{{Ebrahimi} \& {Pharr}(2022)}]{EP22}
{Ebrahimi}, F., \& {Pharr}, M. 2022, \apj, 936, 145

\bibitem[{Frieman \& Rotenberg(1960)}]{frieman-rotenberg1960}
Frieman, E., \& Rotenberg, M. 1960, Reviews of Modern Physics, 32, 898

\bibitem[{{Goedbloed} \& {Keppens}(2022)}]{GK22}
{Goedbloed}, H., \& {Keppens}, R. 2022, \apjs, 259, 65

\bibitem[{Goedbloed {et~al.}(2019)Goedbloed, Keppens, \& Poedts}]{GKP19}
Goedbloed, H., Keppens, R., \& Poedts, S. 2019, Magnetohydrodynamics: Of
  Laboratory and Astrophysical Plasmas (Cambridge University Press)

\bibitem[{Goedbloed(2018{\natexlab{a}})}]{goedbloed2018spectralI}
Goedbloed, J. 2018{\natexlab{a}}, Physics of Plasmas, 25

\bibitem[{Goedbloed(2018{\natexlab{b}})}]{goedbloed2018spectralII}
---. 2018{\natexlab{b}}, Physics of Plasmas, 25

\bibitem[{{Goodman} \& {Xu}(1994)}]{goodmanxu94}
{Goodman}, J., \& {Xu}, G. 1994, \apj, 432, 213

\bibitem[{Hameiri(1981)}]{hameiri1981}
Hameiri, E. 1981, Journal of Mathematical Physics, 22, 2080

\bibitem[{Hawley \& Balbus(1991)}]{HB91}
Hawley, J.~F., \& Balbus, S.~A. 1991, \apj, 376, 223

\bibitem[{Hawley \& Balbus(1992)}]{HB92}
---. 1992, \apj, 400, 595

\bibitem[{{Hawley} {et~al.}(1996){Hawley}, {Gammie}, \& {Balbus}}]{hawley1996}
{Hawley}, J.~F., {Gammie}, C.~F., \& {Balbus}, S.~A. 1996, \apj, 464, 690

\bibitem[{{Hirai} {et~al.}(2018){Hirai}, {Katoh}, {Terada}, \&
  {Kawai}}]{hirai2018}
{Hirai}, K., {Katoh}, Y., {Terada}, N., \& {Kawai}, S. 2018, \apj, 853, 174

\bibitem[{Keppens {et~al.}(2002)Keppens, Casse, \& Goedbloed}]{keppens2002}
Keppens, R., Casse, F., \& Goedbloed, J. 2002, \apj, 569, L121

\bibitem[{Kim \& Ostriker(2000)}]{kim_ostriker}
Kim, W.-T., \& Ostriker, E.~C. 2000, \apj, 540, 372

\bibitem[{Kitchatinov \& R{\"u}diger(2010)}]{kitchatinov2010}
Kitchatinov, L., \& R{\"u}diger, G. 2010, \aap, 513, L1

\bibitem[{Latter {et~al.}(2015)Latter, Fromang, \& Faure}]{latter15}
Latter, H.~N., Fromang, S., \& Faure, J. 2015, \mnras, 453, 3257

\bibitem[{Lehoucq {et~al.}(1998)Lehoucq, Sorensen, \& Yang}]{arpack}
Lehoucq, R.~B., Sorensen, D.~C., \& Yang, C. 1998, ARPACK users' guide:
  solution of large-scale eigenvalue problems with implicitly restarted Arnoldi
  methods (SIAM)

\bibitem[{{Lesaffre} {et~al.}(2009){Lesaffre}, {Balbus}, \&
  {Latter}}]{lesaffre2009}
{Lesaffre}, P., {Balbus}, S.~A., \& {Latter}, H. 2009, \mnras, 396, 779

\bibitem[{Mamatsashvili {et~al.}(2013)Mamatsashvili, Chagelishvili, Bodo, \&
  Rossi}]{mamatsashvili13}
Mamatsashvili, G., Chagelishvili, G., Bodo, G., \& Rossi, P. 2013, \mnras, 435,
  2552

\bibitem[{Matsumoto \& Tajima(1995)}]{matsumoto_tajima95}
Matsumoto, R., \& Tajima, T. 1995, Magnetic viscosity by localized shear flow
  instability in magnetized accretion disks, Tech. rep., Texas Univ., Austin,
  TX (United States). Inst. for Fusion Studies

\bibitem[{{Mishra} {et~al.}(2020){Mishra}, {Begelman}, {Armitage}, \&
  {Simon}}]{mishra2019}
{Mishra}, B., {Begelman}, M.~C., {Armitage}, P.~J., \& {Simon}, J.~B. 2020,
  \mnras, 492, 1855

\bibitem[{Moffatt(1978)}]{moffatt1978}
Moffatt, H.~K. 1978, Cambridge University Press, Cambridge, London, New York,
  Melbourne, 2, 5

\bibitem[{{Noguchi} {et~al.}(2000){Noguchi}, {Tajima}, \&
  {Matsumoto}}]{noguchi2000}
{Noguchi}, K., {Tajima}, T., \& {Matsumoto}, R. 2000, \apj, 541, 802

\bibitem[{Ogilvie \& Pringle(1996)}]{ogilvie_pringle}
Ogilvie, G., \& Pringle, J. 1996, \mnras, 279, 152

\bibitem[{Papaloizou \& Pringle(1984)}]{papaloizou-pringle}
Papaloizou, J., \& Pringle, J. 1984, \mnras, 208, 721

\bibitem[{Pessah \& Psaltis(2005)}]{PP05}
Pessah, M.~E., \& Psaltis, D. 2005, \apj, 628, 879

\bibitem[{Pringle(1981)}]{pringle1981}
Pringle, J.~E. 1981, \araa, 19, 137

\bibitem[{Ripperda {et~al.}(2022)Ripperda, Liska, Chatterjee, Musoke,
  Philippov, Markoff, Tchekhovskoy, \& Younsi}]{ripperda2022}
Ripperda, B., Liska, M., Chatterjee, K., {et~al.} 2022, \apjl, 924, L32

\bibitem[{{Seilmayer} {et~al.}(2014){Seilmayer}, {Galindo}, {Gerbeth},
  {Gundrum}, {Stefani}, {Gellert}, {R{\"u}diger}, {Schultz}, \&
  {Hollerbach}}]{seilmayer2014}
{Seilmayer}, M., {Galindo}, V., {Gerbeth}, G., {et~al.} 2014, \prl, 113, 024505

\bibitem[{Shakura \& Sunyaev(1973)}]{shakura1973}
Shakura, N.~I., \& Sunyaev, R.~A. 1973, \aap, 24, 337

\bibitem[{Spruit {et~al.}(1987)Spruit, Matsuda, Inoue, \& Sawada}]{spruit1987}
Spruit, H., Matsuda, T., Inoue, M., \& Sawada, K. 1987, \mnras, 229, 517

\bibitem[{Terquem \& Papaloizou(1996)}]{terquem_papaloizou}
Terquem, C., \& Papaloizou, J.~C. 1996, \mnras, 279, 767

\bibitem[{van~der Swaluw {et~al.}(2005)van~der Swaluw, Blokland, \&
  Keppens}]{van_der_swaluw05}
van~der Swaluw, E., Blokland, J., \& Keppens, R. 2005, \aap, 444

\bibitem[{Velikhov(1959)}]{velikhov1959}
Velikhov, E. 1959, Sov. Phys. JETP, 36, 995

\bibitem[{Wang {et~al.}(2022{\natexlab{a}})Wang, Gilson, Ebrahimi, Goodman,
  Caspary, Winarto, \& Ji}]{wang2022non-axisymm}
Wang, Y., Gilson, E.~P., Ebrahimi, F., {et~al.} 2022{\natexlab{a}}, Nature
  communications, 13, 4679

\bibitem[{Wang {et~al.}(2022{\natexlab{b}})Wang, Gilson, Ebrahimi, Goodman, \&
  Ji}]{wang2022sMRI}
Wang, Y., Gilson, E.~P., Ebrahimi, F., Goodman, J., \& Ji, H.
  2022{\natexlab{b}}, \prl, 129, 115001

\end{thebibliography}

\end{document}